\documentclass[fleqn,usenatbib,referee]{mnras}

\usepackage{subfig}
\usepackage{float}
\usepackage{footnote}
\usepackage{color,soul}

\makesavenoteenv{tabular}

\usepackage[T1]{fontenc}
\usepackage[utf8]{inputenc}
\usepackage{ae,aecompl}
\usepackage{gensymb}
\DeclareRobustCommand{\VAN}[3]{#2}
\let\VANthebibliography\thebibliography
\def\thebibliography{\DeclareRobustCommand{\VAN}[3]{##3}\VANthebibliography}

\usepackage{graphicx}	
\usepackage{amsmath,bm}	
\usepackage{capt-of} 
\usepackage{multirow}
\usepackage{newtxtext,newtxmath}

\newcommand{\citeg}[1]{\citep[e.g.,][]{#1}}

\title[Timing noise in persistent accreting pulsars]{Pulse Frequency Fluctuations of Persistent Accretion Powered Pulsars}

\author[Serim D. et al.]{
D. Serim,$^{1}$\thanks{E-mail: danjela@astroa.physics.metu.edu.tr}
M. M. Serim,$^{1,2,3}$
A. Baykal$^{1}$
\\
$^{1}$Department of Physics, Middle East Technical University, 06800 Ankara, Turkey\\
$^{2}$Department of Electrical and Electronics Engineering, At{\i}l{\i}m University, 06830 Ankara, Turkey\\
$^{3}$Institut für Astronomie und Astrophysik, Universität Tübingen, 72076, Tübingen, Germany
}

\date{Accepted XXX. Received YYY; in original form ZZZ}

\pubyear{2022}

\begin{document}
\label{firstpage}
\pagerange{\pageref{firstpage}--\pageref{lastpage}}
\maketitle

\begin{abstract}
In this study, we examine the long-term torque noise fluctuations of persistent X-ray binaries Her X-1, Vela X-1, GX 301-2, Cen X-3, 4U 1538-53, OAO 1657-415, and 4U 1626-67 using the historical pulse frequency measurements provided by \emph{CGRO}/BATSE and \emph{Fermi}/GBM. We find that known disk-fed sources exhibit a $1/\omega^2$ red noise component in their power density spectra which is saturated over long timescales. On the other hand, wind-fed sources form a clear white noise component, and the wind-fed sources with occasional transient disk formation imprint $1/\omega$ type flicker noise. We discuss their long-term timing noise properties based on the models to describe the power density spectrum of pulse frequency derivative fluctuations in terms of monochromatic and colored noise processes. Furthermore, we investigate the relation between measured timing noise strengths and other independently measured physical parameters. Despite the low number of sample sources, we suggest that the noise strengths of these sources are correlated with their luminosities and uncorrelated with their magnetic field strengths, implying that the dominant noise-generating mechanism is accretion.
\end{abstract}



\newpage

\section{Introduction}
The study of pulsar timing is beneficial in many areas of astrophysics, and its applications are growing daily. In the early years of their discovery, pulsar timing analysis was used to investigate phenomena like the preciseness of their rotation and the physics of their pulsed emission; however, in recent years, timing analysis has gained popularity and is used in many other crucial areas of astrophysics, such as detection of gravitational waves \citep{2017Abbott}, tests of the general theory of relativity \citep{2003Stairs}, determination of the precise mass of neutron star and the study of ultra-dense matter (see, e.g., \citealt{2016Ozel} and references therein) among others. In principle, the newly emerging areas of astrophysics study the timing irregularities of a source and conclude accordingly. Thus, despite their precise rotational motions, pulsars have sporadic irregularities in their rotation, and these irregularities are classified into two main groups: glitches and timing noise. Glitches are sudden jumps in the spin frequency, and until very recently, they were observed only in isolated pulsars. Nevertheless, \cite{2017Serim} reported a glitch from the pulsar SXP 1062, marking it the first observable glitch detected in a binary system. 

Apart from the glitches, timing irregularities that are classified as timing noise are referred to as the stochastic wandering in the residuals of a particular set of time of arrivals (TOAs) after fitting the data with an appropriate model. Very soon after their discovery, pulsars were observed to have random walks in either phase (PN), frequency (FN) or spin-down rate (SN) \citep{1972Boynton}, and such noise structures can be probed through their power density spectra (i.e., time-dependent noise amplitude distribution). Throughout the years, there have been many studies regarding the timing noise of particular groups of pulsars, either via statistical correlations of timing noise amplitudes with other physical parameters in a selected pulsar subclass or through the investigation of power density spectra of individual sources \citep{1975Groth,1980Cordes2,1994Arzoumanian,2010Shannon,2019Namkham,2020Goncharov,2019Parthasarathy,2020Lower,2020Parthasarathy}. For instance, \cite{1980Cordes1} studied the timing noise of 50 isolated pulsars for the first time; \cite{2010Hobbs} studied the timing noise of 366 radio pulsars for up to 35 years of data; \cite{2010Shannon} studied the timing noise of millisecond pulsars; \cite{2019CerriSerim} studied the timing noise of magnetars, and the timing noise of the pulsars in the binary systems was studied by \cite{1993Baykal}. On the other hand, detailed power spectra of several accreting pulsars are investigated comprehensively by \cite{1997Bildsten} with BATSE data. 

Accretion-powered pulsars reside in binary systems, and their emission is empowered through accreted material from companion stars within the system (see Section \ref{sources} for the general properties of each system studied in this paper). The accretion onto the pulsars can occur via Roche lobe overflow (RLO), wind accretion, or a  combination of these two mechanisms, depending on the type and characteristics of the companion star. The frequency evolutions of neutron stars residing in binary systems are dominated by accretion torques. They are usually much more variable than isolated pulsars due to inhomogeneities in the accretion flow. The individual sources that are the focus of this study are persistent systems characterized by their perpetual emission in X-rays. Since their discovery in the early 1970s, these sources have been the subject of continuous and frequent monitoring due to their persistent emission, which has allowed for an extensive and lengthy set of frequency measurements, mainly maintained by \emph{Fermi}/GBM and \emph{BATSE} which makes such sources good candidates for examining the timing noise on different timescales.

In this work, we present a comprehensive study of the timing noise power density spectra of seven persistent accretion-powered pulsars using the comprehensive data set from \emph{CGRO}/BATSE, and \emph{Fermi/GBM} combined, allowing us to examine the timing noise behavior at longer timescales. Furthermore, we also investigate possible correlations of the timing noise amplitudes with other physical parameters, such as $B$-field and luminosity, and then compare the results with timing noise amplitudes of magnetars \citep{2019CerriSerim} and radio pulsars \citep{2010Hobbs} in similar timescales, aiming to comprehend their noise behavior in a broader range in terms of pulsar population.

\section{Databases}

To examine the pulse frequency fluctuations of persistent X-ray binaries, we utilize the existing frequency measurements in the literature, which are provided by \emph{Fermi}/GBM\footnote{https://gammaray.nsstc.nasa.gov/gbm/science/pulsars.html} and \emph{CGRO}/BATSE\footnote{https://gammaray.nsstc.nasa.gov/batse/pulsar/} monitoring programs. Besides the transient systems, the monitoring campaigns also continuously trace eight persistent systems: Her X-1, Vela X-1, GX 301-2, Cen X-3, 4U 1538-53, GX 1+4, OAO 1657-415, and 4U 1626-67. In this study, we examine all of these sources except GX 1+4, which is comprehensively studied in terms of time-dependent noise strengths by \cite{2017SerimB}. For the remaining seven pulsars, we compile the spin frequency histories generated via measurements of \emph{Fermi}/GBM and \emph{CGRO}/BATSE to achieve the longest possible time interval of input data sets, which will allow us to estimate the noise strengths at longer timescales.

Due to the time gap between the \emph{CGRO}/BATSE and \emph{Fermi}/GBM monitoring programs, the frequency histories for most of the sources span a period of ${\sim}10000$ days with a ${\sim}3000$ days pause in between. For Her X-1, the number of frequency measurements provided by \emph{CGRO}/BATSE is very limited. Thus, we exclude those measurements from our analysis. However, we gather an earlier set of measurements provided in literature \citep{1989Nagase}. For the timing noise strength calculation at the most extended timescale, we utilize the entire data set consisting of both \emph{Fermi}/GBM and \emph{CGRO}/BATSE data (see Section \ref{analysis}), or the frequency data set collected from the literature (in the case of  Her X-1). However, for the shorter timescales, we use only \emph{Fermi}/GBM data set, which provides more precise spin frequency measurements. 

\section{Timing Noise Analysis}
\label{analysis}

Using all the input data, we proceed towards estimating the timing noise strength for each source on various timescales and investigate their noise characteristics. In the literature, there are several distinct approaches to estimate the amplitude of noise of spin frequency fluctuations \citeg{1994Arzoumanian,1972Boynton,1984Deeter,1985Cordes,2010Hobbs,2020Lower}. In this study, we utilize the rms-value technique to calculate the noise strengths of the pulsars \citep{1972Boynton,1984Deeter,1985Cordes}. In this technique, the rms values of the residuals $\langle\sigma_R(m,T)\rangle$ after removal of a polynomial function of order $m$ on a timescale $T$ is connected to the noise strength $S_r$ of order $r$ as 
\begin{center}
\begin{equation}
S_r = \dfrac{\langle\sigma^2_R(m,T)\rangle}{T^{2r-1}\langle\sigma^2_R(m,1)\rangle}_u
\label{sreq}
\end{equation}
\end{center}
where $ \langle\sigma^2_R(m,1)\rangle_u $ is the normalization factor of unit timescale ($T=1$) for unit timing noise strength ($S_r=1$). 
As indicated through Equation \ref{sreq}, the rms values of residuals depend on two parameters: the degree of the polynomial $m$ that is used to describe the regular rotational motion of the pulsar and the time span of the input data set $T$. On the other hand, the normalization factor of the expected rms values can be determined by direct calculations \citep{1984Deeter} or through simulations \citep{2003Scott}. To describe the regular spin evolution of the selected pulsars, we use a quadratic polynomial fit (i.e., $m=2$) for the input data series, which removes up to the $\ddot{\nu}$ component and presume the remaining residuals as noise component. The extracted residuals of the fit are used to estimate the timing noise amplitude for the maximum timescale $T_{max}$ which is limited by the length of the input data, which consists of both \emph{CGRO}/BATSE and \emph{Fermi}/GBM measurements. The same procedure is repeatedly applied for shorter timescales by halving the data set in each case (i.e., $T_{max}/2$, $T_{max}/4$ ...) until the resultant noise amplitude matches with the instrumental noise level\footnote{The instrumental timing noise level is the noise strength level that corresponds to the measuremental errors of the input spin frequencies.}. For the timescales except from $T_{max}$, we use only \emph{Fermi}/GBM measurements.  In our work, we directly utilize the corresponding normalization factor coefficients reported by (\cite{1984Deeter}:Table 1). Then, we determine the power density estimate in each timescale by accumulating the logarithmically re-binned noise strength measurements. Finally, the calculated power density estimates are mapped with their corresponding analysis frequencies ($\omega$) to generate the power density spectra of pulse frequency fluctuations. We further check the form of the power density spectra using the residuals after the removal of cubic trends (i.e., $m=3$) in the frequency histories. We find that the noise strength measurements and power density estimations remain stable for cubic trends as well.

\section{Modeling of Power Spectra}
\label{sec4}
The shape of the power density spectrum reflects the type of noise structure of the pulse frequency fluctuations for a particular source. For example, when the power density spectrum is flat and has a white noise structure (i.e., a power density distribution which is independent of analysis frequency), it implies a noise process of uncorrelated events. On the other hand, when the distribution of the noise amplitudes is analysis frequency dependent, the noise of the pulse frequency fluctuations is characterized by a red noise structure which can be induced by correlated torque-noise processes, such as accretion from a disk.

For example, a monochromatic exponential shot-noise process can yield a $1/\omega^{2}$ red noise component \citep{1997Burderi}. In this case, each torque event is assumed to be identical and exponentially decays on a timescale $\tau$. If a time-dependent external torque generated by accretion flow can be described as:
\begin{equation}
I 2\pi\dot{\nu} (t) = N_{\textrm{ext}}(t) = N_{0} e^{-t/\tau}
\label{torqeq}
\end{equation}
where $N_{0}$ is the mean torque amplitude at event start time ($t=0$). If there are many events, each occurring at the time $t_i$ ($i=0,1,2,..$), each of them will alter the spin frequency derivative of the pulsar as $\delta\dot{\nu}$; then the sequential occurrence of such torque events will lead to a total spin frequency derivative change of $\Delta\dot{\nu}$ as:
\begin{equation}
\Delta\dot{\nu} (t)=  \sum_{i} -\frac{\delta L_{i}}{I \tau} \, e^{-\frac{1}{\tau} (t-t_i)} \,\,\, \textrm{for}\,\,\, t>t_i
\end{equation}
where $\delta L_i$ is the angular momentum change during each event. Note that $\delta L_i$ can be both positive or negative, corresponding to the spin-up ($\delta L_{i,p}$)  and spin-down ($\delta L_{i,n}$) cases (Hereafter, we use subscripts $p$ and $n$ for positive and negative events, respectively).  If we assume that each torque event occurs at the same amplitude ($\delta L_{i,p}= \delta L_{p}$ = constant and $\delta L_{i,n}= \delta L_{n}$ = constant), the Fourier transform for  $\Delta\dot{\nu}$ becomes:
\begin{equation}
\begin{split}
F_{p,n}({\omega}) &= \int_{-\infty}^{+\infty} \sum_{i} -\frac{\delta L_{p,n}}{I \tau} \, e^{-\frac{1}{\tau} (t-t_i)} \, e^{-i\omega t} dt\\
&= -\frac{\delta L_{p,n}/I}{1 + i \tau \omega} \sum_i e^{i\omega t_i}.
\end{split}
\end{equation}
where subscripts $p$ and $n$ denote positive and negative torque events separately. Assuming that the events take place randomly during a total time interval $T$ and they adhere to Poisson distribution with a mean value of $R_{p,n}T$ where $R_p$  and $R_n$ are the rate of positive and negative event occurrences. The power density spectrum of the spin frequency derivative fluctuations $P_{\Delta\dot{\nu},p,n}$ can be calculated in these cases as:
\begin{equation}
\begin{split}
P_{\Delta\dot{\nu},p,n} (\omega) &= \textrm{lim}_{T\to\infty} \frac{1}{T} |F({\omega})|^{2} \\
&= \frac{\delta L_{p,n}^{2}/I^2}{1 + \tau^{2}\omega^{2}} \textrm{lim}_{T\to\infty} \frac{1}{T} \bigg\langle \, \bigg| \sum_i e^{i\omega t_i} \bigg|^{2}\bigg\rangle\\
&= \frac{ R_{p,n} \delta L_{p,n}^2/I^2 }{1 + \tau^{2} \omega^{2}}.
\end{split}
\label{f2eqx}
\end{equation}\noindent Thus, assuming that both negative and positive events decay on the same timescale $\tau$, the total contribution of such events will yield a power density spectrum in the form of:
\begin{equation}
\begin{split}
P_{\Delta\dot{\nu}} (\omega) &= P_{\Delta\dot{\nu},p} (\omega)+P_{\Delta\dot{\nu},n} (\omega) \\
&= \bigg(\frac{R_{p} \delta L_p^2+R_{n} \delta L_n^2}{I^2}\bigg)\bigg( \frac{1}{1 + \tau^{2} \omega^{2}}\bigg).
\end{split}
\label{f2eq}
\end{equation}
At the very low frequencies  (i.e. $\tau>> 1/\omega$), the Equation \ref{f2eq} converges to a constant noise strength of $(R_{p} \delta L_p^2+R_{n} \delta L_n^2)/I^2$. At the higher frequencies, it depicts a power spectrum that is evolving as $\propto \omega^{-2}$. 

On the other hand, a single noise process with a constant relaxation timescale is not sufficient to form a flicker noise \citep{1978Press,1997Burderi,2002Milotti}. It requires at least a colored noise process (i.e., varying relaxation timescales) to generate the observed 1/$\omega$ type component in the power spectrum. Assuming that the shot-noise magnitudes and the event rates $R_p$ and $R_n$ remain constant but the timescale $\tau$ is varying in between $\tau_1$ and $\tau_2$, the power density spectrum of the pulse frequency derivative fluctuations takes the form:
\begin{equation}
\begin{split}
P_{\Delta\dot{\nu}} (\omega) &= \frac{1}{\tau_{2} -\tau_{1}} \int_{\tau_{2}}^{\tau_{1}} \bigg(\frac{R_{p} \delta L_p^2+R_{n} \delta L_n^2}{I^2}\bigg)\bigg( \frac{1}{1 + \tau^{2} \omega^{2}}\bigg) d\tau \\
\\
&=\bigg(\frac{R_{p} \delta L_p^2+R_{n} \delta L_n^2}{I^2}\bigg) \frac{1}{\tau_{2} -\tau_{1}}\,\, \frac{1}{\omega} \bigg[ \tan^{-1}\bigg(\frac{1}{\tau_2 \omega}\bigg) - \tan^{-1}\bigg(\frac{1}{\tau_1 \omega}\bigg)\bigg].
\end{split}
\label{f1eq}
\end{equation}
Using the expansion of the arctan series, the power density spectrum in different timescales can be approximated as:
\begin{equation}
P_{\Delta\dot{\nu}} (\omega) \approx
\begin{cases} \frac{R_{p} \delta L_p^2+R_{n} \delta L_n^2}{I^2}  &\textrm{when} \,\,\, \omega<< 1/\tau_2 << 1/\tau_1 \\
\\
 \frac{R_{p} \delta L_p^2+R_{n} \delta L_n^2}{2 I^2 \omega (\tau_{2} -\tau_{1})} &\textrm{when} \,\,\, 1/\tau_2 << \omega << 1/\tau_2 \\
 \\
 \frac{R_{p} \delta L_p^2+R_{n} \delta L_n^2}{I^2 \omega^2} &\textrm{when} \,\,\, \omega>> 1/\tau_1 >> 1/\tau_2 \\
\end{cases}
\label{f1model}
\end{equation}
The model described through Equation \ref{f1model} converges to the same noise strength $(R_{p} \delta L_p^2+R_{n} \delta L_n^2)/I^2$ found in Equation \ref{f2eq} for low analysis frequencies. However, for the analysis frequencies within range $[\tau_1^{-1}, \tau_2^{-1}]$, it renders a flicker noise component. For the very high analysis frequencies, Equation \ref{f1model} foresees $\omega^{-2}$ proportionality. Unfortunately, this region is generally dominated by instrumental noise (or with an additional white noise component in some cases) in the power spectra illustrated in Section \ref{sources}, and consequently, it is not distinguishable in our analysis. 

For an arbitrary power law index $\Gamma$, the continuum of power density spectra of the colored exponential shot can be generalized as \citep{1997Burderi}:
\begin{equation}
P_{\Delta\dot{\nu}} (\omega) \approx \bigg(\frac{R_{p} \delta L_p^2+R_{n} \delta L_n^2}{I^2}\bigg)\bigg( \frac{1}{1 + (\tau \omega)^{\Gamma}}\bigg).
\label{fgeneral}
\end{equation}
Equations \ref{f2eq}, \ref{f1model} and \ref{fgeneral} provide empirical descriptions for the power density spectra at longer timescales. Therefore, depending on the steepness of red noise components and the general time-dependent noise structure, we use these models to describe the power spectra of the analyzed sources. It should be noted that these empirical models imply a break ($\omega_{break}=1/\tau$) in the power spectrum continuum: a red noise component that converges to a constant noise strength value at low frequencies.
Hence, we only use the power law model to describe the power density spectrum if the break frequency is not observable. Furthermore, we model the white noise structures (i.e., when $\Gamma\simeq0$) that appear in the power spectra with a constant noise level $S_r$. 
The resulting best-fit parameters after modeling individual spectra are presented in Table \ref{tab2}. 
\begin{table}
\begin{center}
\caption{The model parameters for power density spectra of the analyzed sources.}
\label{tab2}
\begin{tabular}{ l c c c c c }
Source Name& Model & $S_r$& $\Gamma$ & $\omega_{break}$& Fit range\\
\hline
Vela X-1	&	White	&	$	3.00	\pm	0.40	\times	10^{-20}	$		&	$	-			$	&	$	-					$	&	$	1.03	\times	10^{-9}	-	2	\times	10^{-7}	$	\\
4U 1538-52	&	White	&	$	6.26	\pm	0.64	\times	10^{-21}	$		&	$	-			$	&	$	-					$	&	$	1.14	\times	10^{-9}	-	1	\times	10^{-7}	$	\\
4U 1626-67	&	Power Law	&	$	4.30	\pm	2.00	\times	10^{-49}	$		&	$	-3.5	\pm	1.2	$	&	$	-					$	&	$	1.03	\times	10^{-9}	-	6.1	\times	10^{-9}	$	\\
OAO 1657-415	&	Equation 8	&	$	2.59	\pm	0.26	\times	10^{-17}	$		&	$	\bm{-1} 			$	&	$	1.25	\pm	0.19	\times	10^{-8}	$	&	$	1.15	\times	10^{-9}	-	3.6	\times	10^{-7}	$	\\
		&	Power Law	&	$	1.07	\pm	0.80	\times	10^{-23}	$		&	$	-0.79	\pm	0.06	$	&	$	-					$	&	$	3	\times	10^{-9}	-	3.6	\times	10^{-7}	$	\\
GX 301-2	&	Equation 8 +WN	&	$	1.20	\pm	0.23	\times	10^{-18}	$		&	$	\bm{-1}			$	&		$\bm{8\times 10^{-9}}$, $\bm{5\times 10^{-8}}$		&	$	1	\times	10^{-9}	-	3.8	\times	10^{-7}	$	\\
		&	Power Law	&	$	2.95	\pm	2.82	\times	10^{-23}	$		&	$	-0.54	\pm	0.06	$	&	$	-	$	&	$	1	\times	10^{-9}	-	5	\times	10^{-8}	$	\\
Her X-1	&	Equation 6	&	$	2.10	\pm	0.40	\times	10^{-19}	$		&	$	\bm{-2}			$	&	$	6.4	\pm	1.60	\times	10^{-8}	$	&	$	6.2	\times	10^{-10}	-	2.8	\times	10^{-7}	$	\\
		&	Power Law	&	$	3.60	\pm	2.60	\times	10^{-26}	$		&	$	-0.91	\pm	0.08	$	&	$		$	&	$	2.5	\times	10^{-8}	-	1	\times	10^{-7}	$	\\
Cen X-3	&	Equation 6 	&	$	1.46	\pm	0.13	\times	10^{-17}	$		&	$	\bm{-2}			$	&	$	7.15	\pm	0.84	\times	10^{-8}	$	&	$	6	\times	10^{-9}	-	4	\times	10^{-7}	$	\\
		&	Power Law	&	$	4.50	\pm	3.29	\times	10^{-26}	$		&	$	-1.16	\pm	0.16	$	&	$	-					$	&	$	5	\times	10^{-8}	-	4	\times	10^{-7}	$	\\

\end{tabular}
\end{center}
\noindent Notes: \\
1) The parameters indicated in bold are kept frozen during model fitting.\\
2) $S_r$ values correspond to the coefficient $(R_{p} \delta L_p^2+R_{n} \delta L_n^2)/I^2$ in the aforementioned equations in Section \protect\ref{sec4}.

\end{table}
\section{Selected Sources and Their Noise Spectra}
\label{sources}

\subsection{4U 1626-67}
{4U 1626-67} is an ultra-compact LMXB system composed of a pulsar and a very-low mass companion star ($< 0.1 M_{\odot}$), called Kz TrA \citep{1971Giacconi,1977McClintock}. The accreting pulsar has a spin-period of $\sim$7.68 s, a rapid orbital period of only 42 min, and an estimated distance of  $3.5^{+2.3}_{-1.3}$ kpc \citep{1981Middleditch,1998Chakrabarty,2019Schulz,2020Malacaria}. The detailed spectral studies of 4U 1626-67 with \emph{BeppoSAX} reveals the existence of cyclotron resonant scattering feature (CRSF) around ${\sim}37$ keV, which leads to an inferred dipole magnetic field strength of $3.2 \times \, 10^{12}$ G \citep{1998Orlandini}. Owing to its persistent emission, the 4U 1626-67 has been monitored numerously since its discovery, and the source has undergone two major torque reversals, first in 1990 \citep{1993Wilson} and later in 2008 \citep{2020Malacaria}. During these torque reversals, the source switches from spin-up to spin-down or vice versa with the frequency derivative magnitude in the order of $10^{-13}$ Hz s$^{-1}$ \citep{1997Chakrabarty,2010CameroArranz}. 
\begin{figure}
\includegraphics[width=0.5\columnwidth]{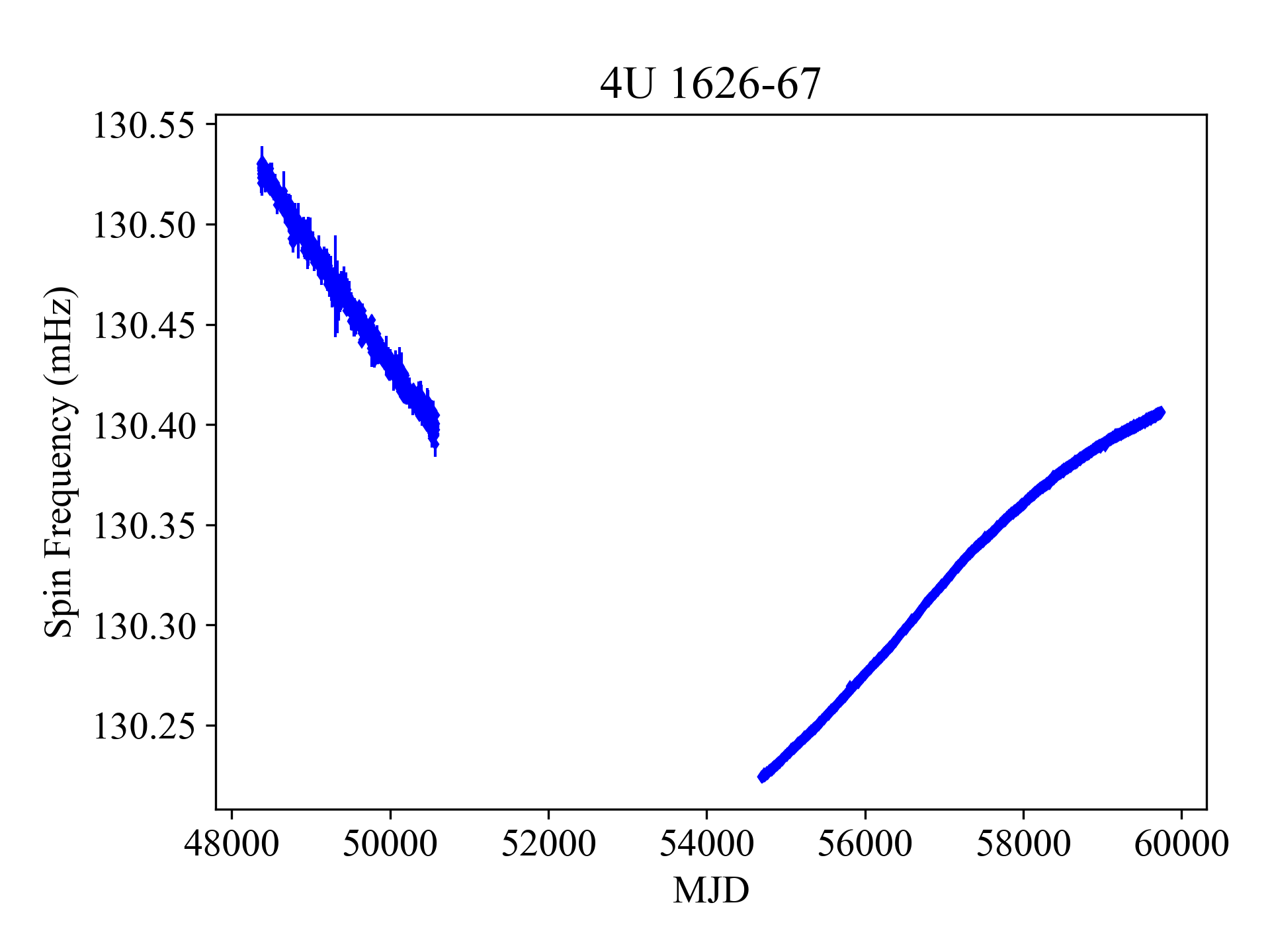}
\includegraphics[width=0.5\columnwidth]{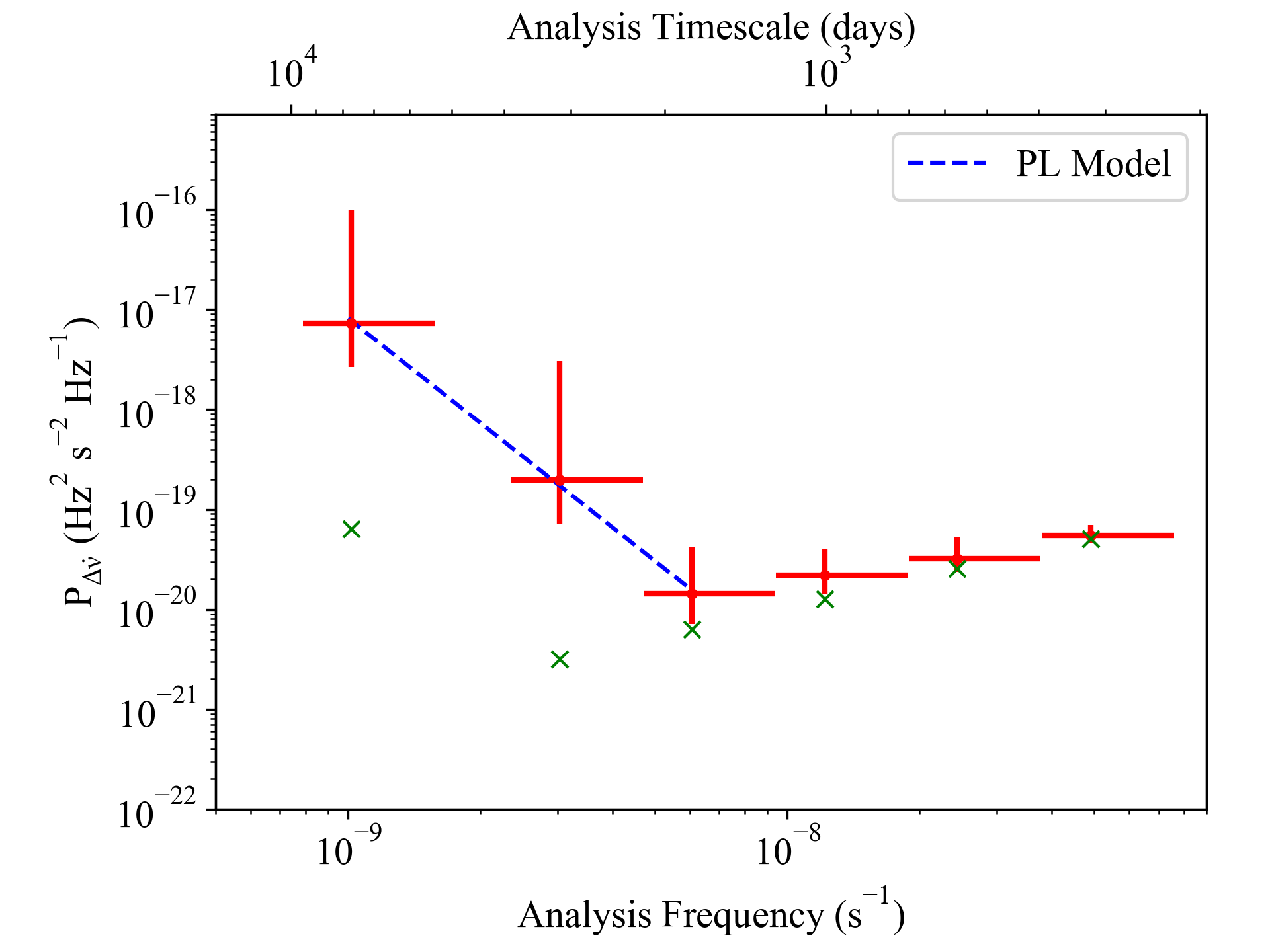}
\captionof{figure}{Left panel: The spin frequency data set of 4U 1626-67 used for noise strength analysis. Right panel: The power density spectrum of 4U 1626-67. Red marks indicate the power density estimates at corresponding analysis frequencies. Green crosses demonstrate the instrumental noise level at given analysis frequencies. The blue dashed line represents the best fit of the power law model in the analysis frequency range of $	1.03	\times	10^{-9}	-	6.1	\times	10^{-9}	$ s$^{-1}$ }
\label{4u1626m67noise}
\end{figure}

The constructed power density spectrum (Figure \ref{4u1626m67noise}) shows that 4U 1626-67 is characterized by a bimodal noise structure. In the timescales shorter than ${\sim}2000$ days (or $\omega\gtrsim6\times10^{-9}$ s$^{-1}$), the power density estimates are scaling up with instrumental noise level without a signature of a white noise component. On the other hand, in the timescales longer than ${\sim}2000$ days, the power density spectrum exhibits a red noise component. The power density spectrum of  4U 1626-67 does not exhibit a low-frequency break. Thus, we describe its power spectral continuum with a power law model, which yields a power law index of $\Gamma = -3.5\pm 1.2$. Throughout the analysis frequency range of the power spectrum, the noise strength of 4U 1626-67 varies in between ${\sim}2.9 \times 10^{-18}$ to ${\sim}1.4 \times 10^{-20}$ Hz$^{2}$ s$^{-2}$ Hz$^{-1}$.

\subsection{GX 301-2}
{GX 301-2} is a HMXB system discovered in 1976 when \emph{Ariel 5} satellite detected the ${\sim}680$ s pulsations of its pulsar  \citep{1976White}. The companion star is suggested to be the B1-type hypergiant, Wray 77 \citep{1980Parkes}, whose mass loss via stellar wind is estimated to be of the order of ${\sim}10^{-5}\, \rm{M_{\odot}} $/yr \citep{2006Kaper}.
The orbital period of the pulsar is relatively long ($P_{orb} \sim 41.5$ days), with high eccentricity ($e \sim 0.5$), and its X-ray flux alters depending on the orbital phase \citep{2010Doroshenko}. The inferred magnetic field of the pulsar is $ \sim 3 \times 10^{12} $ G \citep{2004Kreykenbohm} and the \emph{Gaia} estimated distance is $3.5^{+0.6}_{-0.5} $ kpc \citep{2020Malacaria}. Early observations with \emph{CGRO}/BATSE showed that  GX 301-2 pulsar is characterized by a relatively stable rotational motion with small spin frequency fluctuations possibly originating from wind accretion \citep{1997Koh}. Nevertheless, GX 301-2 is also characterized by several relatively long (30-40 days) spin-up episodes, which are linked with a possible transition disk \citep{1997Koh,2019Nabizadeh}.

\begin{figure}
\includegraphics[width=0.5\columnwidth]{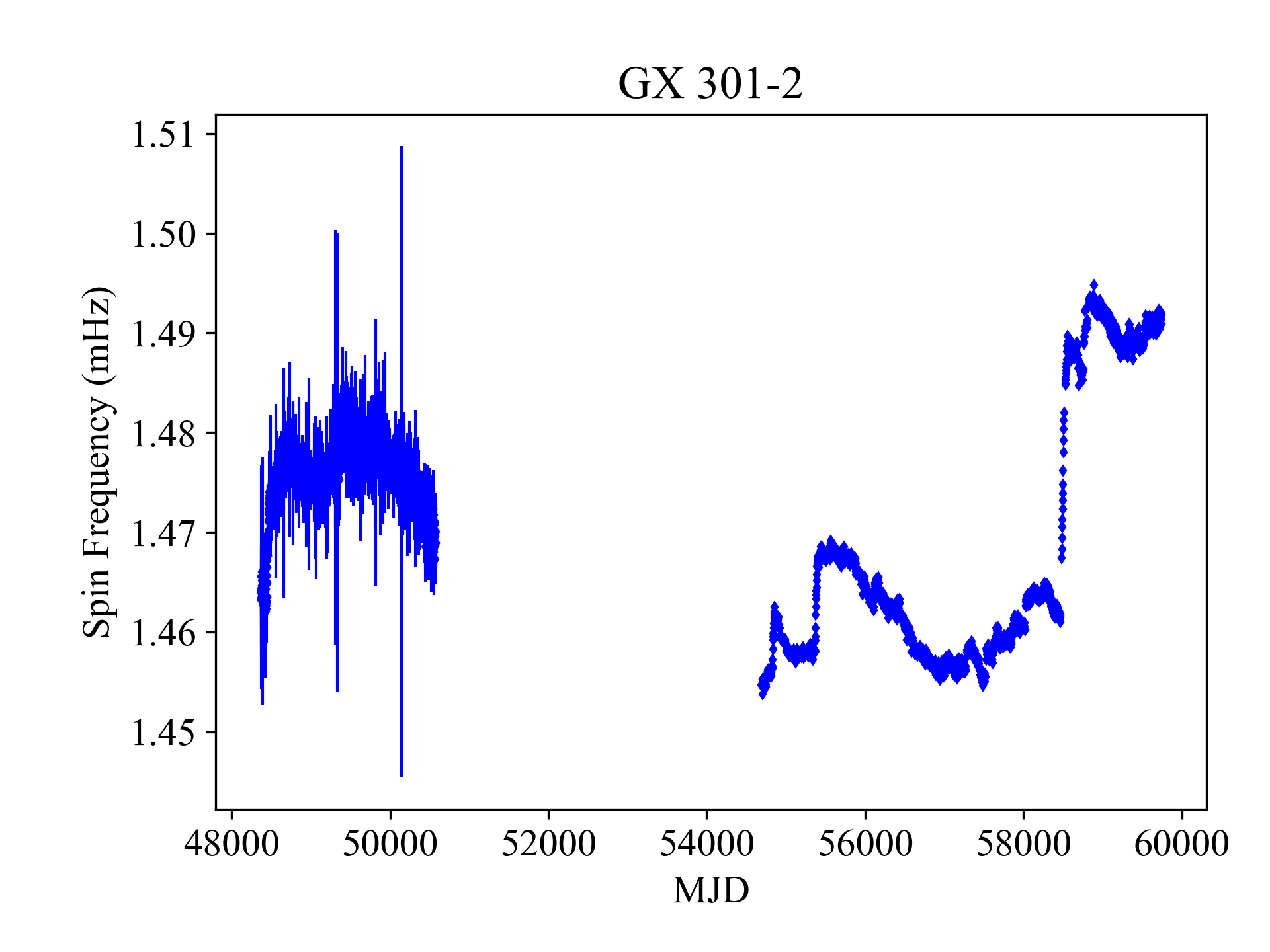}
\includegraphics[width=0.5\columnwidth]{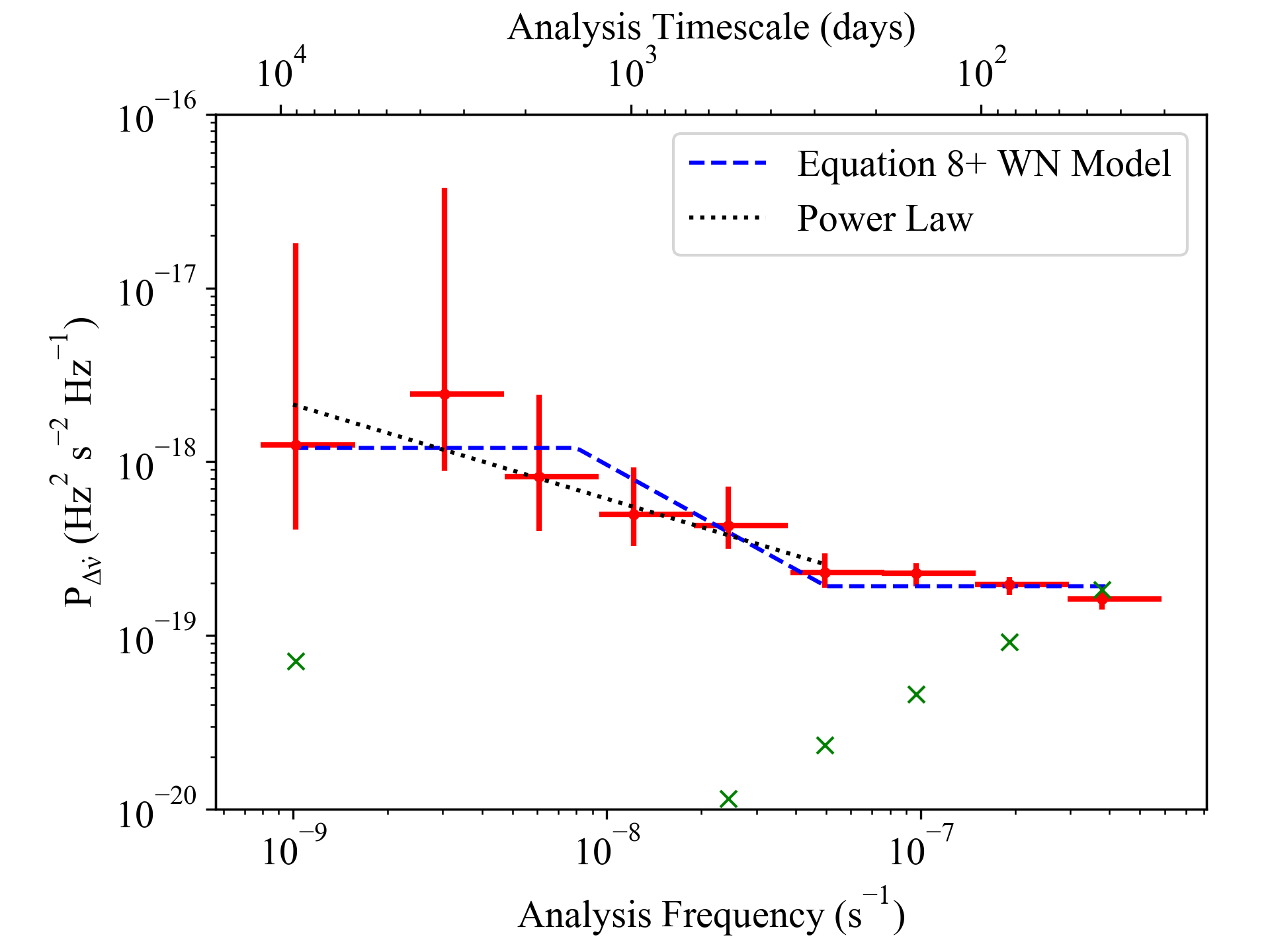}
\caption{Left panel: the spin frequency data set of GX 301-2 used for noise strength analysis. Right panel: The power density spectrum of GX 301-2. Red marks indicate the power density estimates at corresponding analysis frequencies. Green crosses demonstrate the instrumental noise level at given analysis frequencies.  The black dotted line represents the best fit of the power law model in the analysis frequency range of $	1	\times	10^{-9}	-	5	\times	10^{-8}	$ s$^{-1}$. The blue dashed line indicates the best fit of Equation \protect\ref{f1model} which is modified with a white noise model at high frequencies (see text for details).}
\label{gx301m2noise}
\end{figure}

The power density spectrum of GX 301-2 is shown in Figure \ref{gx301m2noise}. The overall power density estimates for GX 301-2 resides in a range of ${\sim}2.5 \times 10^{-19}$ to ${\sim}2.5\times 10^{-18}$ Hz$^{2}$ s$^{-2}$ Hz$^{-1}$. The power density spectrum of GX 301-2 is characterized by a white noise structure in the short timescale measurements up to ${\simeq}250$ days ($\omega$ $\simeq 4.5\times 10^{-8}$ s$^{-1}$), which switches to a red noise structure ($\propto \omega^{-1}$) in the longer time intervals ($>250$ days).  The steepness of the red noise component also tends to diminish at lower frequencies ($\omega<7 \times 10^{-9}$ s$^{-1}$); however, considering the low resolution of the power density estimates at these analysis frequencies, we cannot ascertain the break frequency of the power density spectrum (if there is any). Consequently, we use two separate models to represent the power spectra. In the first model, we use Equation \ref{f1model} to describe the power density estimates at low analysis frequencies, which is further modified with a constant to demonstrate the dominant white noise structure at high analysis frequencies. Taking the equivocacy of the break analysis frequencies into account, we freeze the break frequencies as $8\times 10^{-9}$ and $5\times 10^{-8}$ s$^{-1}$ during the fitting. In the second model, we use a simple power law model for the analysis frequency range $1\times10^{-9}-	5	\times	10^{-8}	$  s$^{-1}$ where the red noise component is present. In this case, the power law index $\Gamma$ is obtained as $-0.54\pm 0.06$. 

\subsection{4U 1538-52}

{4U 1538-52} is a HMXB system that exhibits relatively slow pulsations with a period of ${\sim}527$ s \citep{1977Davison}. Its companion is suggested to be a B0Iab-type supergiant,  QV Nor \citep{1978Parkes}. The orbital period of the system is found to be ${\sim}3.7$ days, which decays at a rate of  $\dot{P}_{orb}/P_{orb}=(0.4 \pm 1.8) \times 10^{-6}\, \rm{yr^{-1}} $ \citep{2006Baykal}. \emph{Gaia} measured distance of the system is $6.6^{+2.2}_{-1.5}$ kpc \citep{2018BailerJones}, and the source is reported to have a CRSF at ${\sim}23$ keV in its X-ray spectrum \citep{1990Clark}. The estimated magnetic field of 4U 1538-52 is ${\sim}2 \times 10^{12}$ G \citep{2014Hemphill}. 

\begin{figure}
\includegraphics[width=0.5\columnwidth]{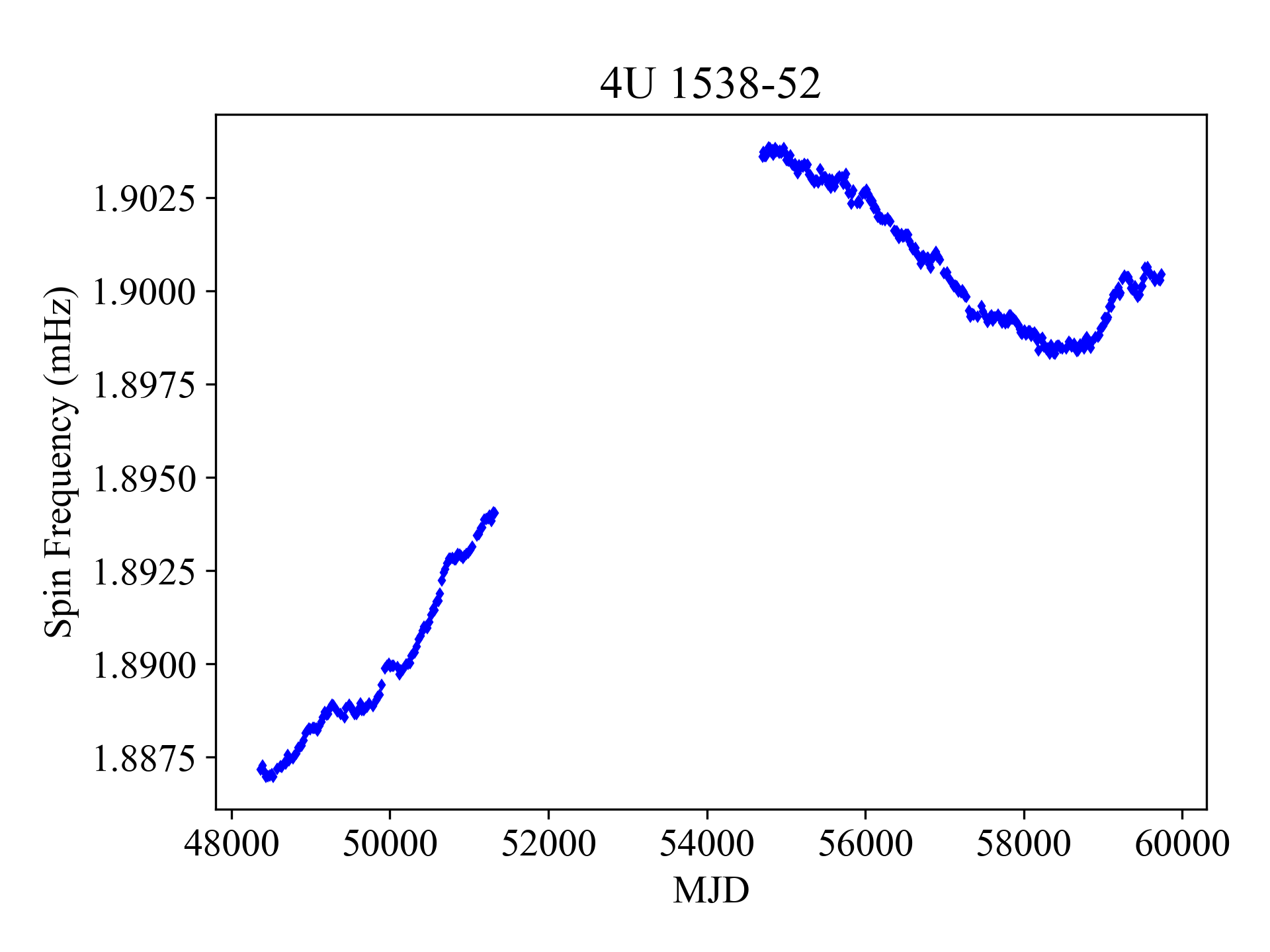}
\includegraphics[width=0.5\columnwidth]{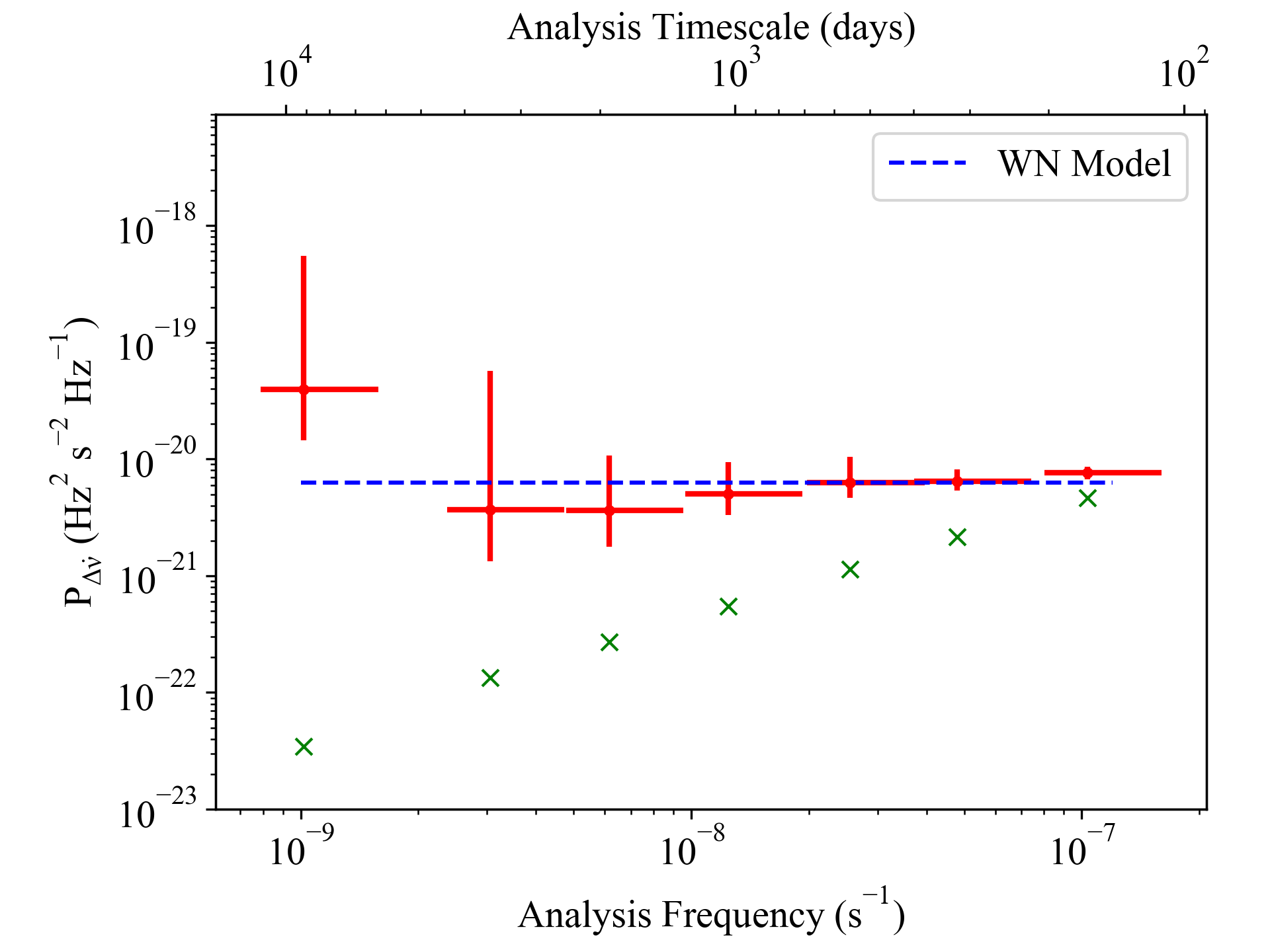}
\caption{Left panel: The spin frequency data set of 4U 1538-52 used for noise strength analysis. Right panel: The power density spectrum of 4U 1538-52. Red marks indicate the power density estimates at corresponding analysis frequencies. Green crosses demonstrate the instrumental noise level at given analysis frequencies. The blue dashed line indicates the best fit of a white noise model.}
\label{4u1538m52 noise}
\end{figure}

The power density spectrum of the pulse frequency derivative fluctuations of 4U 1538-52 (Figure  \ref{4u1538m52 noise}) indicates that the noise strength remains constant around ${\sim}4{-}8 \times 10^{-21}$ Hz$^{2}$ s$^{-2}$ Hz$^{-1}$ level and the source has a white noise structure throughout almost all the analysis frequency ranges. The only exception is the excess noise measurement reaching up to ${\sim}5 \times 10^{-20}$ Hz$^{2}$ s$^{-2}$ Hz$^{-1}$ at the longest timescale. This excess noise level may hint at a formation of a red noise component at lower analysis frequencies. {4U 1538-52} shows super orbital intensity modulations around $\sim14.91$ days, which are thought to be related to $\dot{\nu}$ fluctuations \citep{2021Corbet}. Unfortunately, we cannot observe its consequences because such timescales in our analysis are dominated by measuremental noise. 

\subsection{OAO 1657-415}

{OAO 1657-415} is a HMXB system whose existence was revealed by \cite{1978Polidan} when they detected the pulsar with the \emph{Copernicus} satellite. Following observations with \emph{HEAO A-2} revealed a pulse period of $ P_{s}\, {\sim}38.2\,\rm{s}$ \citep{1979White}. Using the \emph{CGRO}/BATSE data, \cite{1993Chakrabarty} reported an orbital period of  $ P_{\rm{orb}}\, {\sim}10.4 $ days and showed the system is an eclipsing X-ray binary with  $ T_{\rm{eclipse}}\,{\sim}1.7 $ days. The companion of the OAO 1657-415 system is lately identified as an Ofpe/WNL type of star, which is being transformed from a main sequence OB star to a relatively rare type Wolf-Rayet star \citep{2009Mason}. Very recently, \cite{2022Sharma} detected CRSF in the \emph{NuStar} spectra of the source at ${\sim}36$ keV, which translates to a magnetic field strength of ${\sim}3.3 \times 10^{12} $ G. 

The continuous monitoring of the source with \emph{CGRO}/BATSE and Fermi/GBM pulsar project has shown that OAO 1657-415 is characterized by a stochastic spin-up/spin-down evolution with an average spin-up rate of $ \dot{\nu}\,{\sim}8.3 \times 10^{-13}\,\, \rm{Hz}\,\rm{s^{-1}}$ \citep{1997Baykal,1997Bildsten,2000Baykal}. Even though it is a wind-fed system, \cite{2012Jenke} suggest the possibility of two different accretion modes for OAO 1657-415. The first mode is the torque-flux correlated regime in which a stable accretion disk is expected to be present. In the second mode, the torque-flux correlation vanishes.
 \begin{figure}
\includegraphics[width=0.5\columnwidth]{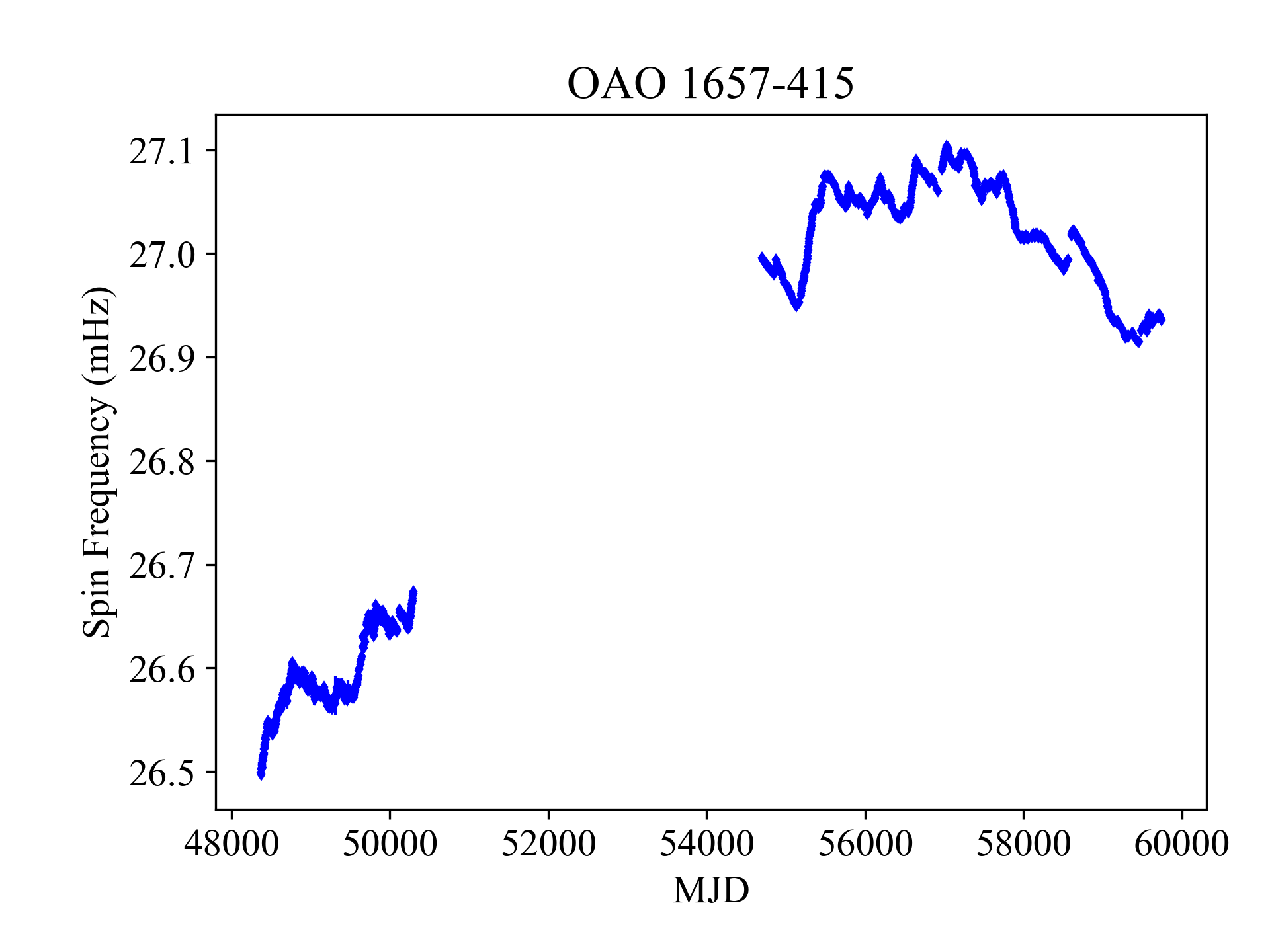}
\includegraphics[width=0.5\columnwidth]{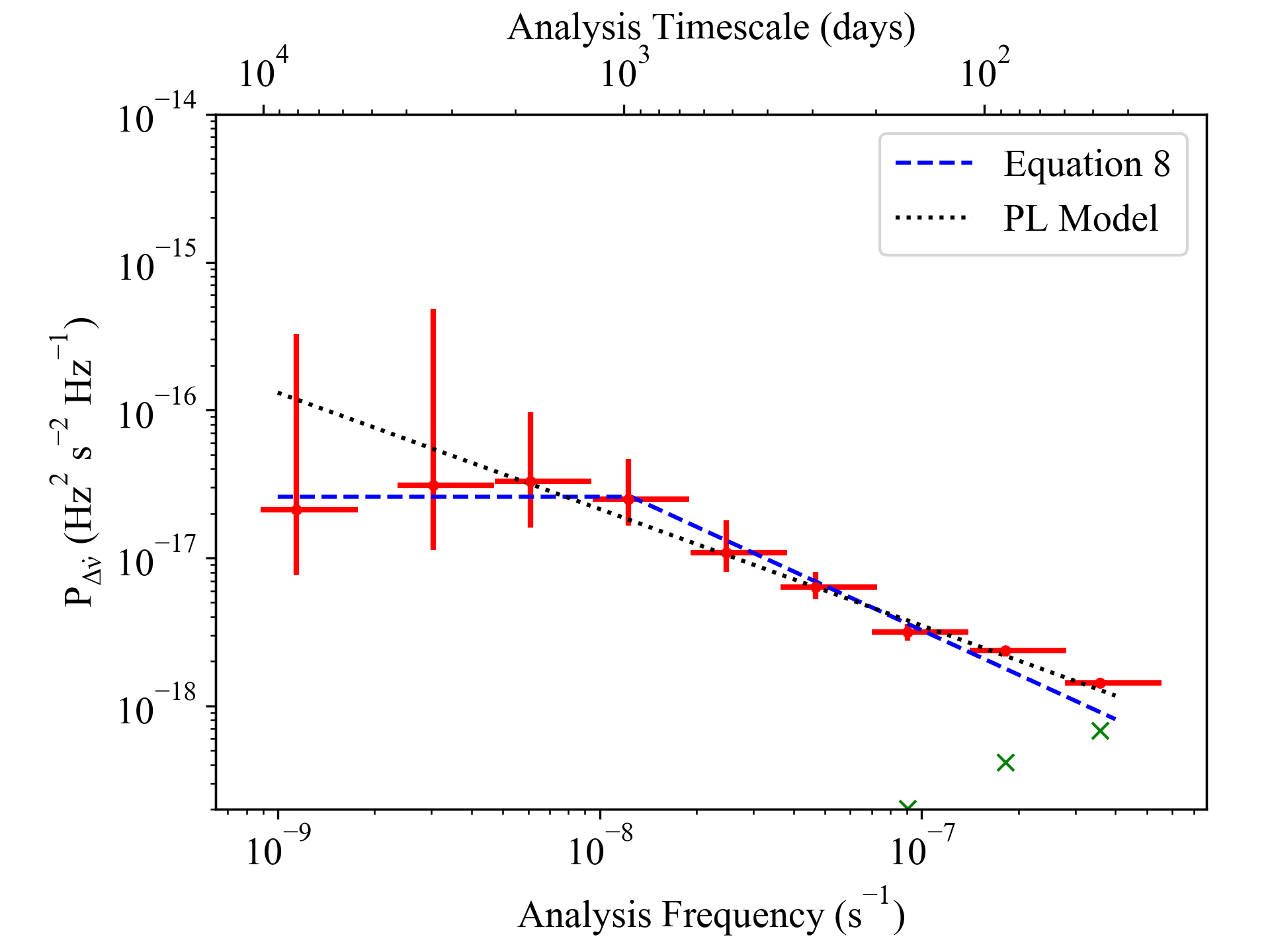}
\caption{Left panel: The spin frequency data set of OAO 1657-415 used for noise strength analysis. Right panel: The power density spectrum of OAO 1657-415. Red marks indicate the power density estimates at corresponding analysis frequencies. Green crosses demonstrate the instrumental noise level at given analysis frequencies. The black dotted line represents the best fit of the power law model in the analysis frequency range of $	3	\times	10^{-9}	-	3.6	\times	10^{-7}	$ s$^{-1}$. The blue dashed line indicates the best fit of Equation \protect\ref{f1model}.}
\label{oao1657m415noise}
\end{figure}

Using the frequency measurements provided by \emph{Fermi}/GBM and \emph{CGRO}/BATSE, we generated the power spectrum for OAO 1657-415. The noise structure of OAO 1657-415 behaves as a flicker noise ($\sim1/\omega$) for almost all the analysis frequencies; however, the power density estimates seem to be flattened on the level of ${\sim}3\times 10^{-17}$ Hz$^{2}$ s$^{-2}$ Hz$^{-1}$ for the analysis frequencies lower than ${\sim}10^{-8}$ s$^{-1}$ (see Figure \ref{oao1657m415noise}). Nonetheless, we cannot rule out the possibility of the red noise component that may persist at low analysis frequencies. Hence, we depict the continuum of the power density spectrum with two cases. In the first case, we used the colored shot-noise model for $1/\omega$ noise specified in Equation \ref{f1model}. In the second case, we use a simple power law model that covers all the analysis frequency ranges. In the first case, fitting Equation \ref{f1model} yields $\omega_{break}=1.25\pm0.19\times10^{-8}$ s$^{-1}$. In the latter case, the power density spectrum can be represented with a power law model whose index is obtained as $\Gamma=-0.79\pm0.06$ within the analysis frequency range of $3\times10^{-9} -3.6\times10^{-7}$ s$^{-1}$.  

 \subsection{Vela X-1}

Vela X-1 is a well-known X-ray binary and one of the most studied. The binary nature of the system was discovered in 1975 when \emph{SAS-3} observatory data revealed ${\sim}283 $  s pulsations from its pulsar \citep{1977McClintock}. The companion is a supergiant B0.5Ia type of star named HD 77581 \citep{1972Hiltner}. The orbit of the system is of low eccentricity  ($ e \sim 0.09$) and has a period of $P_{orb} \sim 8.9\, \rm{days}$ \citep{1995vanKerkwijk}. The pulse frequency history of the pulsar is characterized by random spin-up/spin-down episodes, which is considered to be a sign of wind accretion \citep{1989Deeter}. From the CRSF, the magnetic field of the pulsar is inferred as ${\sim}2.1 \times 10^{12} \,\rm{G}$ \citep{2014bFurst}, and the \emph{Gaia} estimated distance is $2.42^{+0.19}_{-0.17}\, \rm{kpc}$ \citep{2020Malacaria}.

\begin{figure}
\includegraphics[width=0.5\columnwidth]{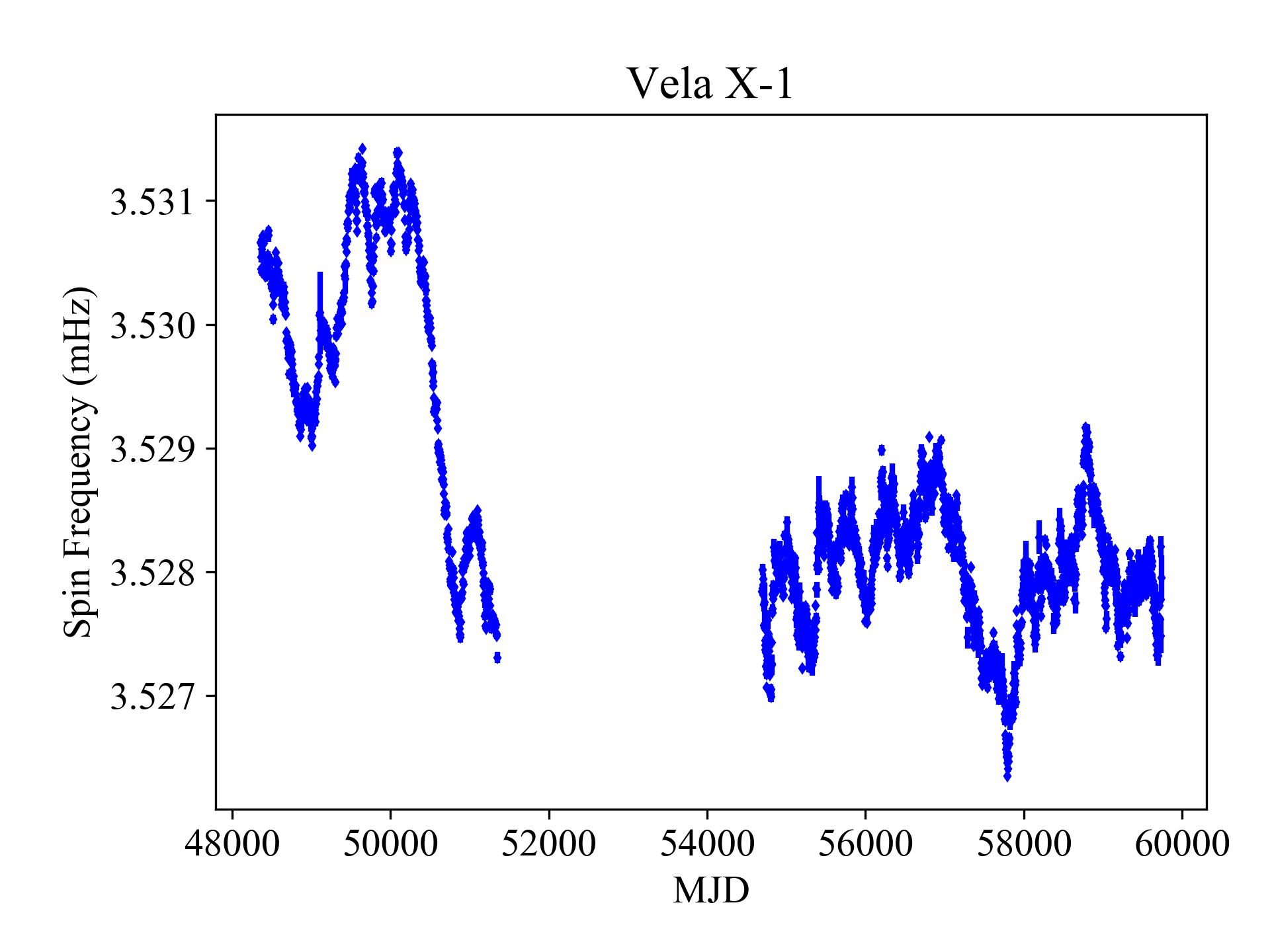}
\includegraphics[width=0.5\linewidth]{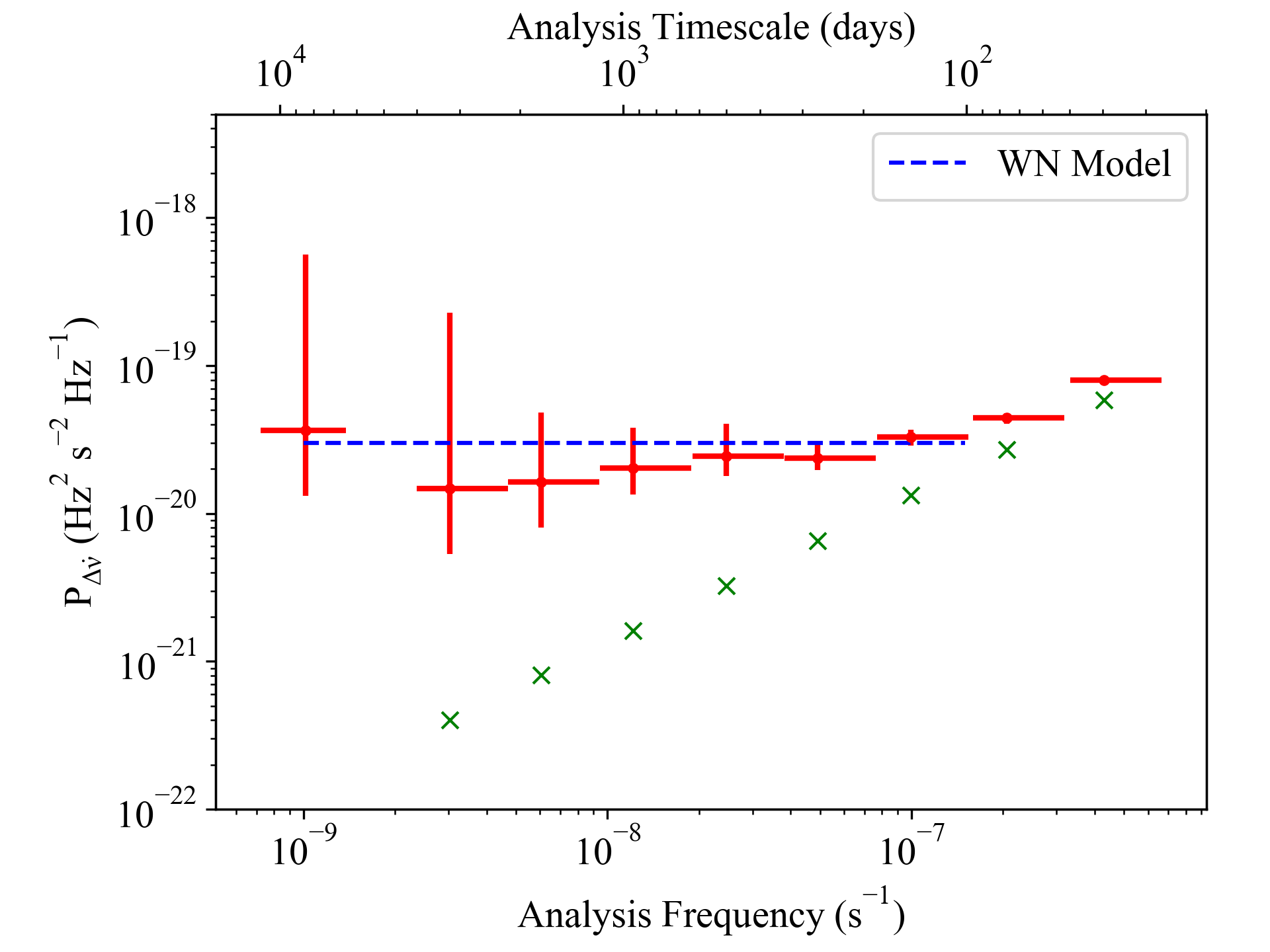}
\caption{Left panel: The spin frequency data set of Vela X-1 used for noise strength analysis. Right panel: The power density spectrum of Vela X-1. Red marks indicate the power density estimates at corresponding analysis frequencies. Green crosses demonstrate the instrumental noise level at given analysis frequencies. The blue dashed line indicates the best fit of a white noise model.}
\label{velax1noise}
\end{figure}

For Vela X-1, the data sampling rate of \emph{Fermi}/GBM measurements allows us to estimate the timing noise strength at small timescales down to ${\sim}23$ days. Throughout all the analysis frequencies, the estimated noise strength amplitudes stay in the range of ${\sim}1.5 - 5 \times 10^{-20}$ Hz$^{2}$ s$^{-2}$ Hz$^{-1}$ (Figure \ref{velax1noise}). The entire power spectrum of Vela X-1 is compatible with a white noise structure expected in wind-fed systems in all the analysis frequency ranges except for the slight excess noise at the longest time scale. The excess noise might be connected to the possible cyclic turnovers in the spin-up/down behavior of Vela X-1 at time scales of 17-19 year \citep{2021Chandra}. 

\subsection{Her X-1}
{Her X-1} is a LMXB system discovered in 1971. \cite{1972Tananbaum} revealed that the spin period $P_s \sim 1.2\, \rm{s}$ of the pulsar using \emph{Uhuru} data. The companion is a ${\sim}2.2 \, \rm{M_{\odot}}$ A7-type star known as HZ Her \citep{1976Middleditch}, and the \emph{Gaia} estimated distance of the system is $5.0^{+0.8}_{-0.6} \, \rm{kpc}$ \citep{2020Malacaria}.
With its low-mass nature, the pulsar accretes via Roche lobe overflow, has an orbital period of $ P_{orb}\,{\sim}1.7\, \rm{days}$, and is characterized by a super orbit X-ray modulation with a period of $ \sim 35\, \rm{days}$ \citep{1972Bahcall,1973Giacconi}. The CRSF of Her X-1 indicates a magnetic field value of $B \sim 3.5 \times 10^{12}\, \rm{G}$ \citep{2007Staubert}. 

\begin{figure}
\includegraphics[width=0.5\columnwidth]{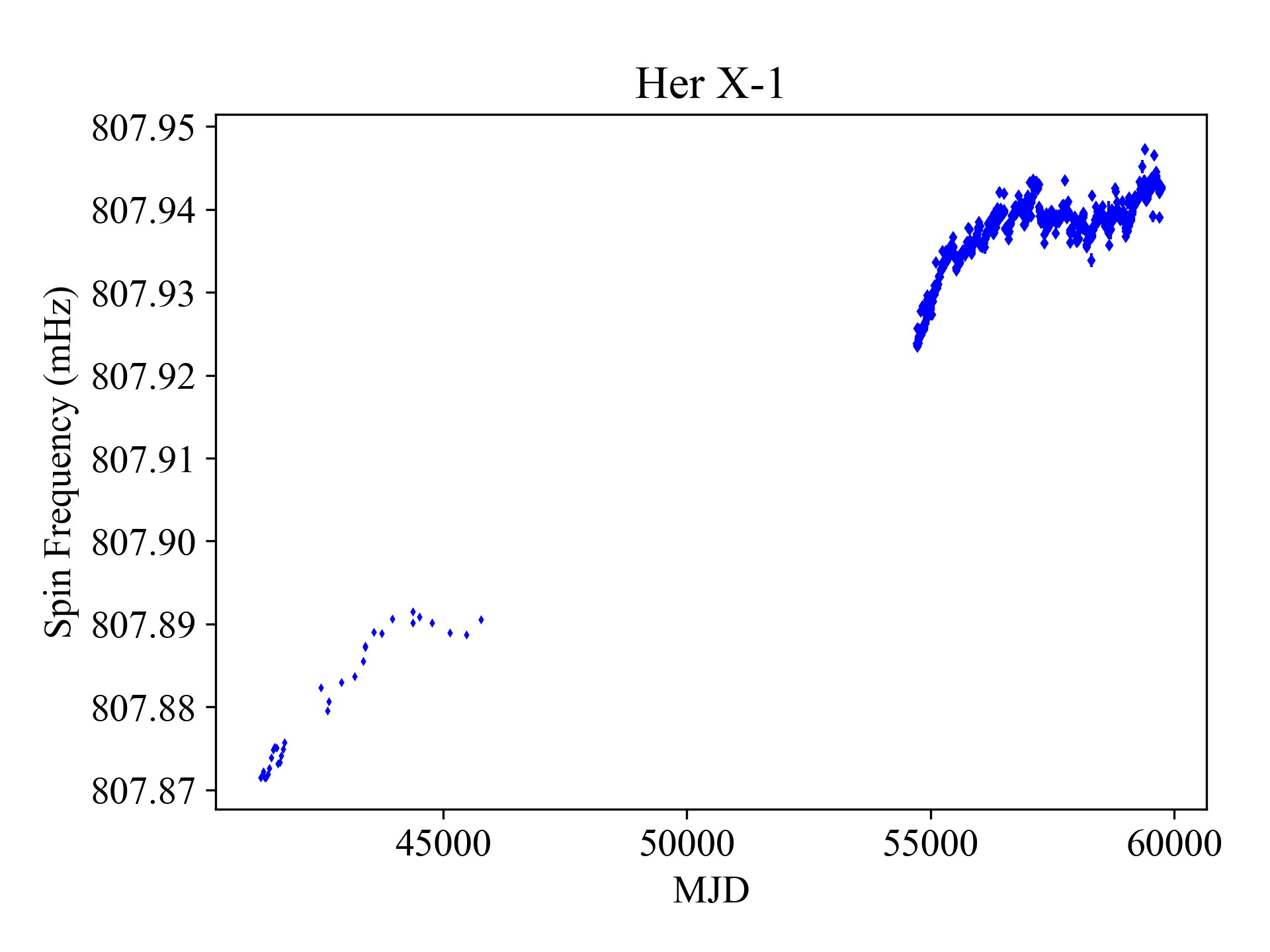}
\includegraphics[width=0.5\linewidth]{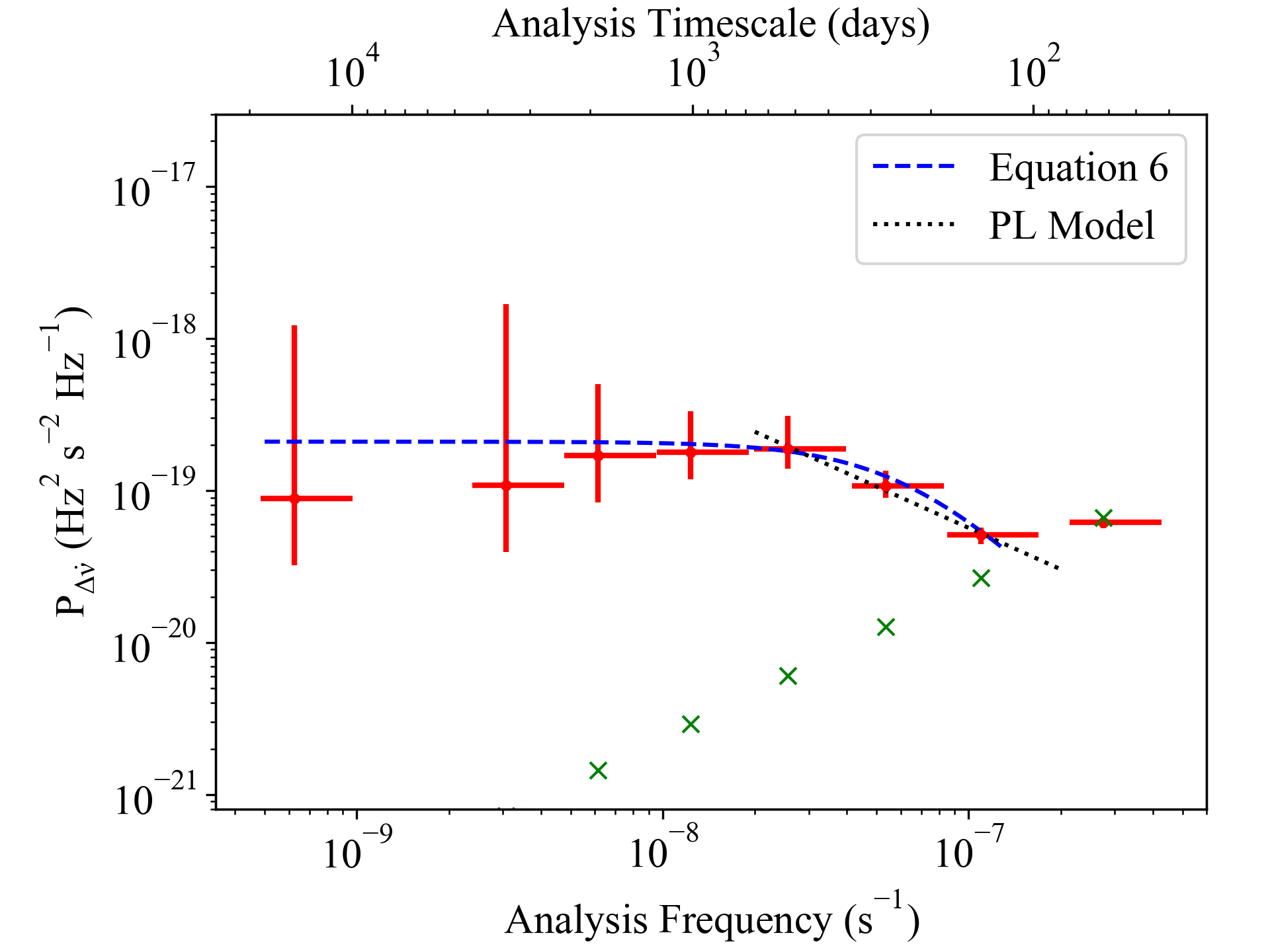}
\caption{Left panel: The spin frequency data set of Her X-1 used for noise strength analysis. Right panel: The power density spectrum of Her X-1. Red marks indicate the power density estimates at corresponding analysis frequencies. Green crosses demonstrate the instrumental noise level at given analysis frequencies. The black dotted line represents the best fit of the power law model in the analysis frequency range of $	3	\times	10^{-8}	-	1	\times	10^{-7}	$ s$^{-1}$. The blue dashed line indicates the best fit of Equation \protect\ref{f2eq}.}
\label{herx1noise}
\end{figure}

In Figure \ref{herx1noise}, we present the power density spectrum of Her X-1. The noise strength estimates for Her X-1 drift from ${\sim}2 \times 10^{-19}$ to ${\sim}5 \times 10^{-20}$ Hz$^{2}$ s$^{-2}$ Hz$^{-1}$ as they approach higher analysis frequencies. At low frequencies, the power spectrum is flat at the level of ${\sim}2 \times 10^{-19}$  Hz$^{2}$ s$^{-2}$ Hz$^{-1}$; however, Her X-1 exhibits a steep red noise component in a narrow analysis frequency interval (${\sim}3 \times 10^{-8}$ s$^{-1}$ -- ${\sim}10^{-7}$ s$^{-1}$). Fitting a simple power law within this range yields a power law index of $\Gamma = -0.91\pm 0.08$. At higher frequencies, the instrumental noise dominates the pulse frequency derivative fluctuations. Therefore, we describe the power density spectrum of Her X-1 with the monochromatic shot-noise model characterized by Equation \ref{f2eq}. In this case, we obtain the low-frequency break as $6.4\pm 1.6 \times 10^{-8}$ s$^{-1}$.

\subsection{Cen X-3}

{Cen X-3} is the first pulsar detected in the X-ray energy band and one of the most studied. Using the \emph{Uhuru} X-ray observatory data, ${\sim}4.8\, \rm{s}$ pulsations of the neutron star were discovered \citep{1971Giacconi}. It orbits a supergiant companion,  V779 Cen \citep{1974Krzeminski}, with an orbital period of $P_{\rm{orb}}\, {\sim}2.1 $ days \citep{1983bKelley}. 
Similar to 4U 1538-52 \citep{2006Baykal}, Cen X-3 is also observed to have an orbital decay of $\dot{P}_{\rm{orb}}/P_{\rm{orb}} \sim -1.8 \times 10^{-6}\, \rm{yr}^{-1}$, probably due to the tidal dissipation and/or mass transfer nature between the two stars of the binary \citep{1989Nagase}. The pulsar main accretion scheme is via RLO with additional contribution from wind accretion \citep{1997Bildsten}. Interpreting the absorption feature around ${\sim}28$ keV in its spectra as CRSF, Cen X-3 is anticipated to have a magnetic field strength of $\rm{B}\, {\sim}2.6 \times 10^{12}$ G \citep{1998Santangelo,2008Suchy}.

\begin{figure}
\includegraphics[width=0.5\columnwidth]{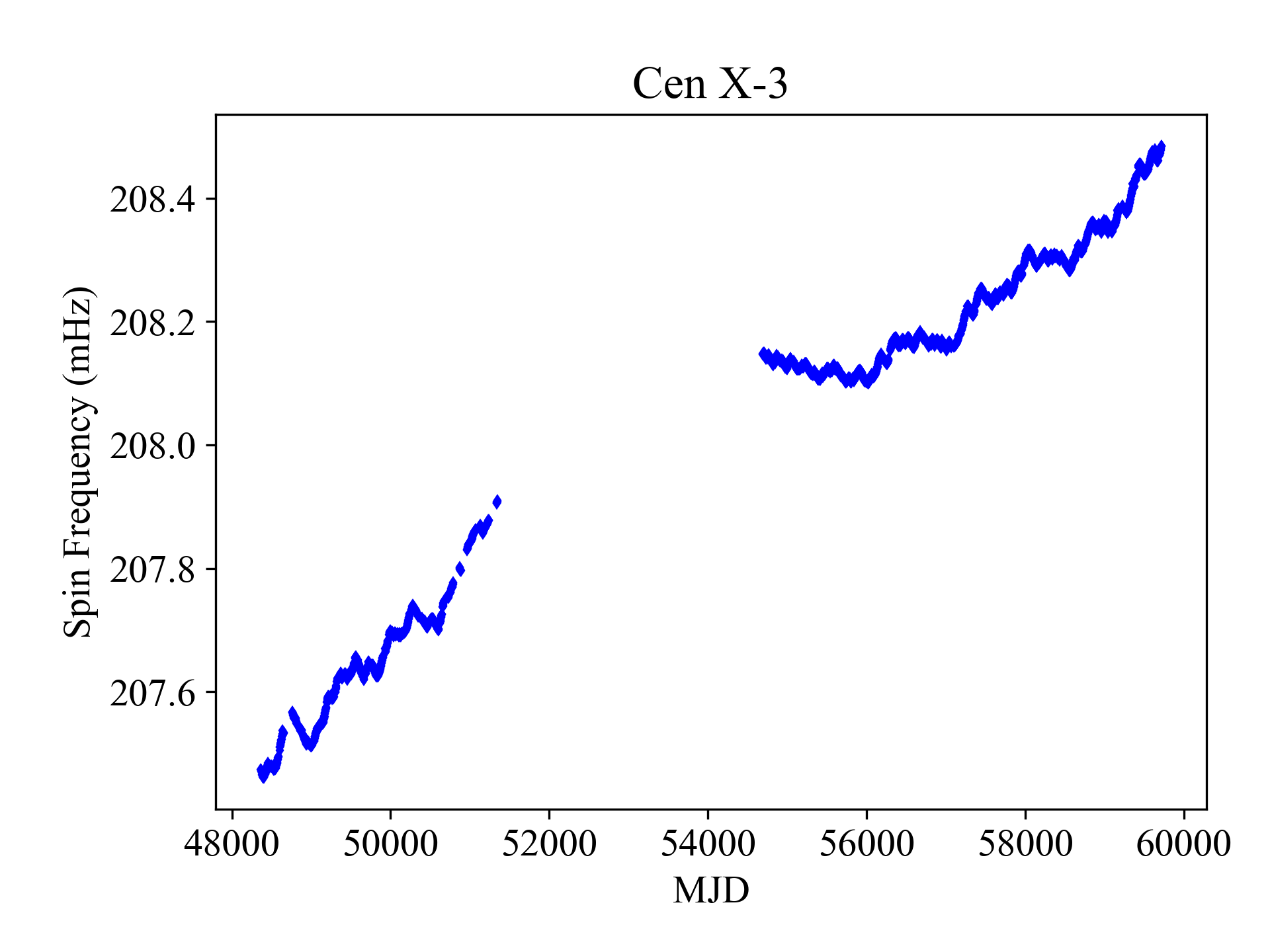}
\includegraphics[width=0.5\linewidth]{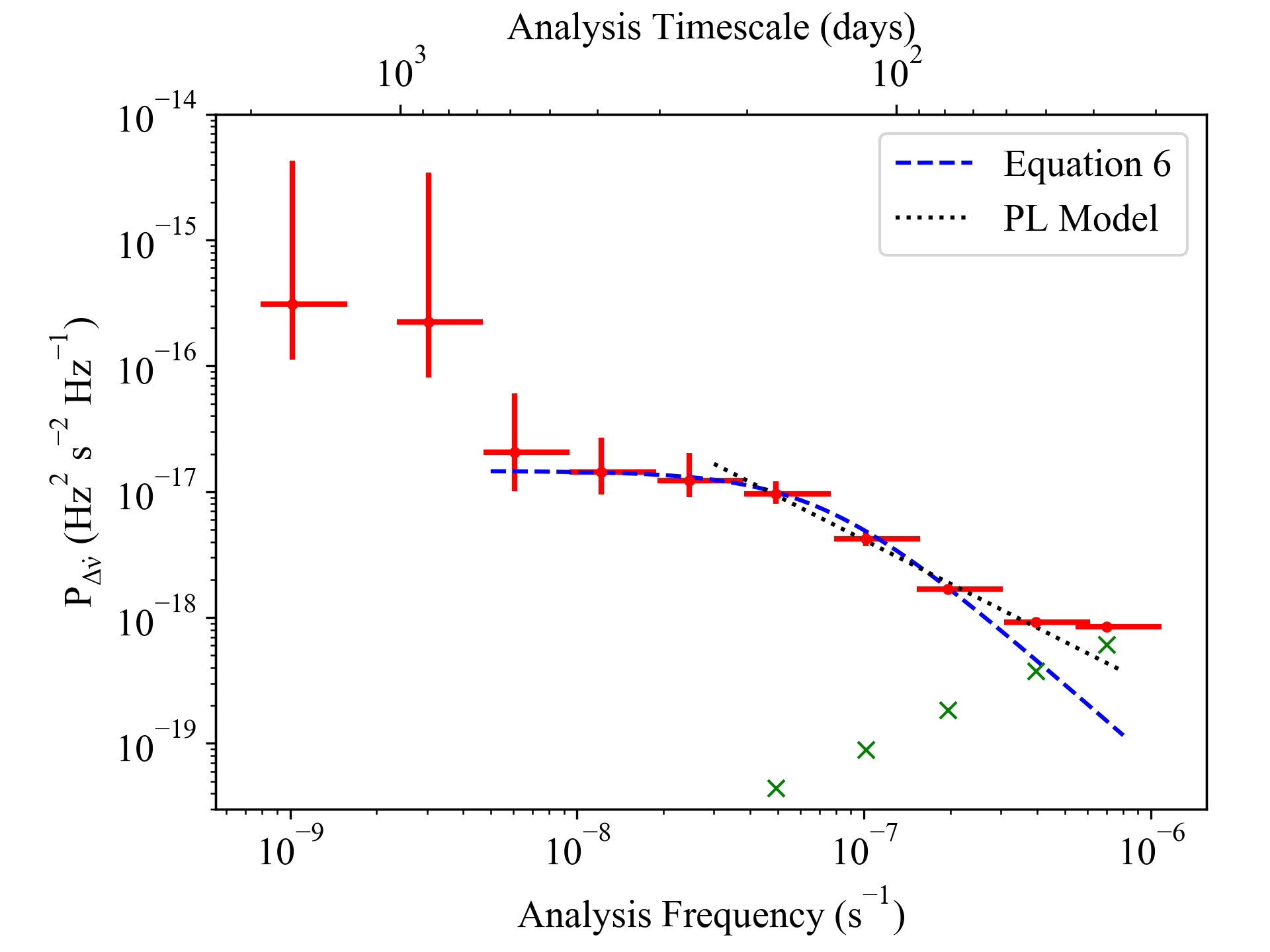}
\caption{Left panel: The spin frequency data set of Cen X-3 used for noise strength analysis. Right panel: The power density spectrum of Cen X-3. Red marks indicate the power density estimates at corresponding analysis frequencies. Green crosses demonstrate the instrumental noise level at given analysis frequencies. The black dotted line represents the best fit of the power law model in the analysis frequency range of $	5	\times	10^{-8}	-	4	\times	10^{-7}	$ s$^{-1}$. The blue dashed line indicates the best fit of Equation \protect\ref{f2eq}.}
\label{cenx3noise}
\end{figure}

The constructed power density spectra of Cen X-3 and Her X-1 are similar, but {Cen X-3} also exhibits two excess noise estimations at very low frequencies. The noise strength estimates vary from ${\sim}2.8 \times 10^{-16}$ to ${\sim}8.4 \times 10^{-19}$ Hz$^{2}$ s$^{-2}$ Hz$^{-1}$.  Apart from the excess noise estimates, Cen X-3 exhibits a white noise structure on the level of $\sim 10^{-17}$ Hz$^{2}$ s$^{-2}$ Hz$^{-1}$ in the analysis frequency range ${\sim}6 \times 10^{-9} - {\sim 5}\times 10^{-8}$ s$^{-1}$. At higher frequencies ($\omega> 5\times10^{-8}$ s$^{-1}$), the spectrum clearly demonstrates the existence of a red noise component until it is dominated by the instrumental noise. When a simple power law model is used for this analysis frequency interval ($5\times 10^{-8}$ to $4\times 10^{-7}$ s$^{-1}$), the resulting power law index becomes $1.16\pm0.16$. Therefore, we proceed with Equation \ref{f2eq} to represent the overall continuum of the power density spectrum except for the two excess noise estimations. In this case, the break analysis frequency is obtained as $7.15\pm0.84\times 10^{-8}$ s$^{-1}$.

\section{Relationship of noise strengths with physical quantities}

We examine the relationship between the timing noise strength values of the studied sources and their physical parameters, such as magnetic field ($B$) and X-ray luminosity\footnote{Given the uncertainties in the bolometric luminosities of the selected sources, we employ a new parameter that is correlated with 15-50 keV X-ray luminosity (See Section \ref{secLS} for details).}. The aim is to understand if there is a relation between noise strengths and other independently measured physical quantities, which may provide hints about the dominant mechanisms causing the fluctuations in the spin frequency derivatives of the studied sources. As it is evident in the results presented in Section \ref{sources}, the timing noise strengths may vary on different timescales, especially when a red noise component is present. Thus, it is crucial to refer to the timing noise strengths at similar timescales to make a consistent comparison across the sample sources. In our study, we use a timescale of $T \sim 1000$ days, allowing us to compare the noise strengths of our source sample with that of magnetars and radio pulsars, which were previously investigated in similar timescales \citep{2019CerriSerim}. 
\begin{table}
\begin{center}
\caption{The list of distances and magnetic field strengths of the sources used in statistical correlations.}
\label{tab1}
\begin{tabular}{ l  c c c }
Source Name&  $B_{12}^{a}$&Distance $^{c}$\\
\hline
4U 1626-67& 3.2 & 3.5$^{+2.3}_{-1.3}$  &  \\
GX 301-2&  3  & 3.5$^{+0.6}_{-0.5}$  &  \\
4U 1538-52&  2 & 6.6$^{+2.2}_{-1.5}$  &  \\
OAO 1657-415 &   3.3$^{b}$& 7.1$^{+1.3}_{-1.3}$   & \\
Vela X-1 & 2.1 & 2.42$^{+0.19}_{-0.17}$  & \\
Her X-1 & 3.5 & 5.0$^{+0.8}_{-0.6}$  & \\
Cen X-3 & 2.6 &6.4$^{+1.4}_{-1.1}$  &\\
GX 1+4&  -& 7.6$^{4.3}_{2.8}$  &\\
\end{tabular}
\end{center}
\noindent $^{(a)}$ The magnetic values are taken from \protect\cite{2015Revnivtsev} and references therein.\\
\noindent $^{(b)}$ Obtained from \protect\cite{2022Sharma}.\\
\noindent $^{(c)}$ The enlisted distances are obtained from \emph{Gaia} (\protect{\cite{2020Malacaria}} and references therein.)\\
\end{table}

\subsection{Magnetic Field vs. Timing Noise}

\begin{figure}
	\includegraphics[width=1\linewidth]{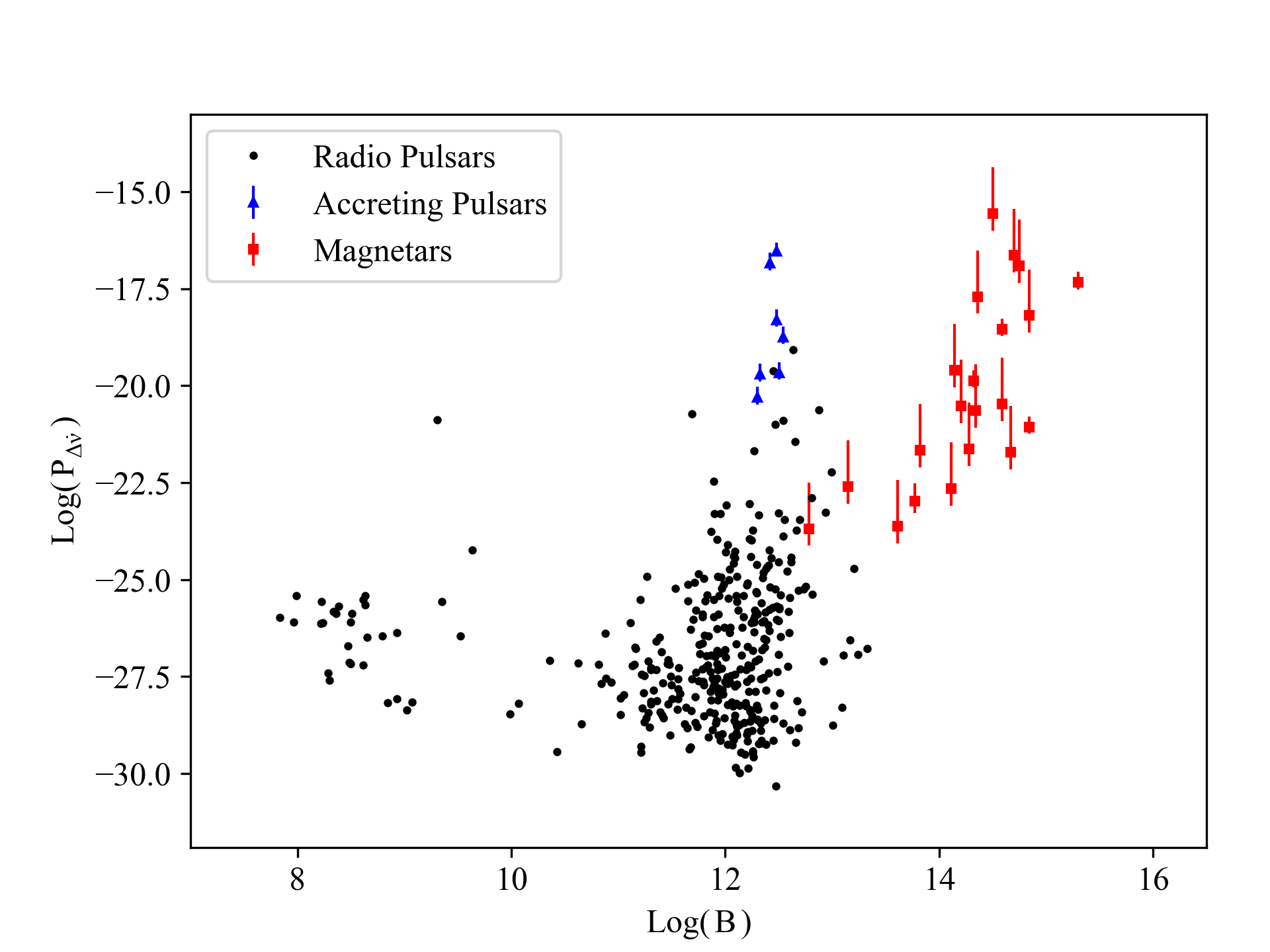}
\caption{The distribution of power density estimates at ${\sim}1000$-day timescales with respect to the magnetic field strengths across different pulsar populations. The noise strengths values for pulsars and magnetars are gathered from \protect\cite{2010Hobbs} and \protect\cite{2019CerriSerim}. The magnetic field strengths for accreting systems, which are deduced from cyclotron features, are complied from \protect\cite{2015Revnivtsev}.}
 \label{BvN}
\end{figure}

We investigate the relationship between the $B$-field and the timing noise strength of our source sample and compare the results with that of isolated pulsars represented in \cite{2019CerriSerim}. Methodologically, the measure of physical parameters of accretion-powered pulsars differs from that of isolated pulsars. For instance, dipolar magnetic field strengths of isolated pulsars are often inferred via magnetic braking (i.e., through the relation $B = 3.2\times 10^{19} \sqrt{P\dot{P}}$ G).  On the other hand, the $B$-field of accreting pulsars is either found by modeling the cyclotron feature in their respective spectrum or approximated via torque models. Since the latter approximation is rather model-dependent and leads to diverse results \citep{2019Staubert}, we proceed with the $B$-field strengths deduced via cyclotron features (see \citep{2015Revnivtsev} and references therein). 
The resulting distribution of $B$-field versus timing noise strengths is illustrated in Figure \ref{BvN}.

In the diagram of $B$-field vs. noise strengths, we observe that persistent accreting pulsars form a narrow correlative clump in the middle upper part of the figure. It can also be seen that the magnetars and persistent systems have similar timing noise levels. Considerable effort has been expended to find debris disks around magnetars over the years. The idea of such debris discs is supported through optical and infrared observations \citep{2001Kaplan,2001Hulleman,2006Wang,2006Ertan,2008Mereghetti,2009Kaplan,2013Trumper}. The distinction between the accreting pulsar and magnetar samples in the $B$-field vs. noise strength diagram may hint that they should have different origins for the torque fluctuations. It should be understood that the findings do not rule out the possibility of fossil disks encircling magnetars; nevertheless, even if they exist, it is possible that they are not the cause of the torque fluctuations.
The low number of accreting pulsar samples with a narrow distribution of the deduced magnetic field strengths may be the reason why the sources examined in this study do not converge to a correlative structure in the B-field vs. noise strength diagram, or the existence of a correlation is rather inconclusive.
On the other hand, it has been demonstrated that the magnetars display a correlative behavior that is related to the fluctuations in the magnetosphere \citep{2019CerriSerim,2013Tsang}. To reach more concrete results, a further investigation with more accreting sources is required.

\subsection{X-ray Luminosity vs. Timing Noise}
\label{secLS}
\begin{figure}
	\includegraphics[width=1\linewidth]{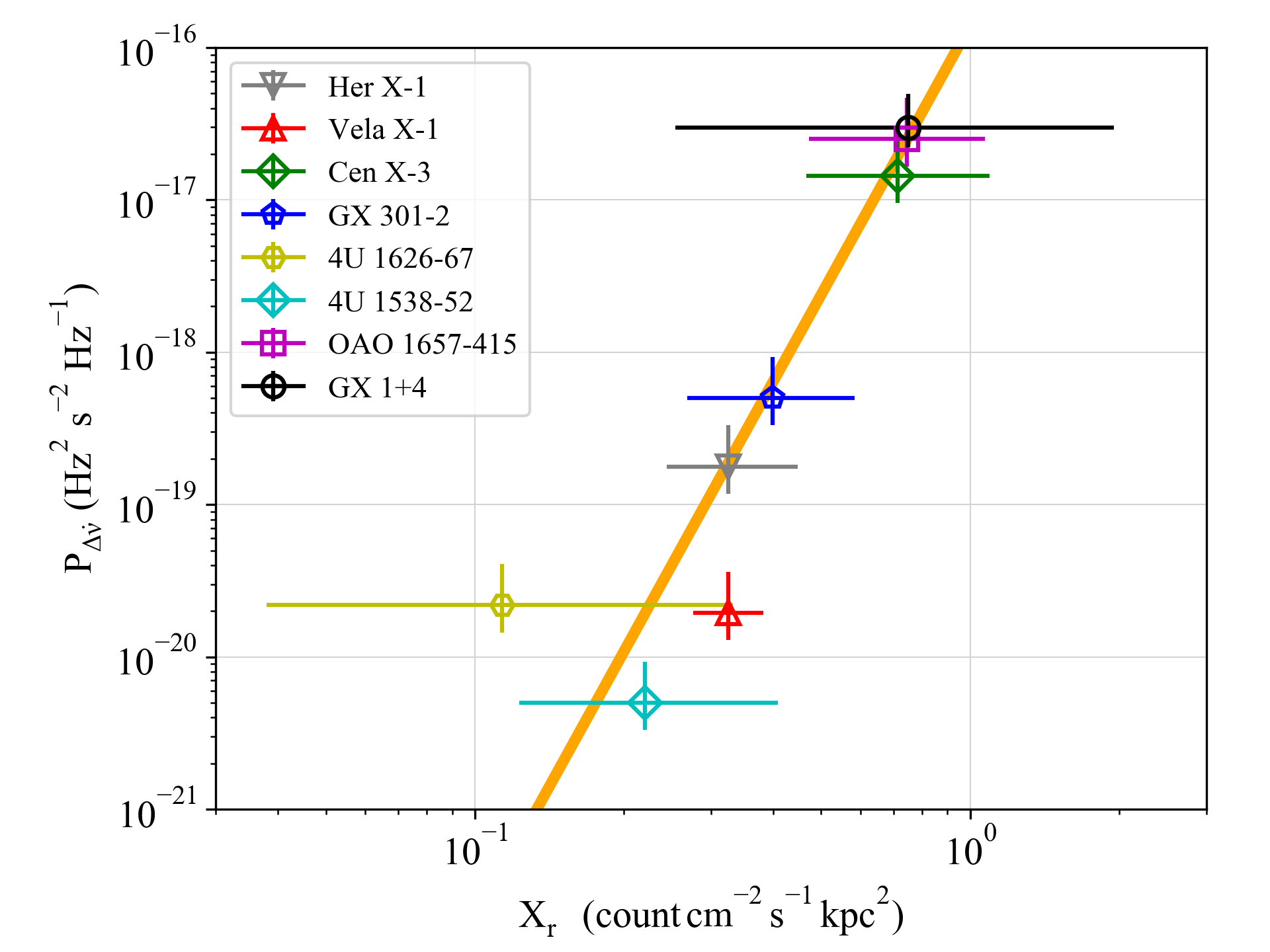}
\caption{The distribution of power density estimates at  ${\sim}1000$-day timescales as a function of rescaled count rates $X_r$. The error range of $X_r$ reflects the uncertainties of the distance measurements. Solid line represents the best fit for the data.}
    \label{LvN}
\end{figure}

According to the torque models, the torque exerted on a pulsar originates from mass accretion, which is converted to X-ray luminosity. However, the X-ray luminosity in such models refers to the bolometric luminosity in the entire X-ray energy band (i.e., ${\sim}1{-}200$ keV). 
In general, the emission in higher energy bands is interpolated by extending the spectral models in the lower energy bands. 
In order to make a meaningful comparison with other sources, the average bolometric X-ray luminosity within the same interval of the input frequency data set is required. However, it is impractical to obtain flux measurements for the entire data set for all the sources. Therefore, we use the following approach to roughly demonstrate the relation between the noise strengths and X-ray luminosities. We first obtain an average 15-50 keV count rate of the \emph{Swift}/BAT measurements\footnote{https://swift.gsfc.nasa.gov/results/transients/} within the time interval of noise strength measurements, assuming that it is a suitable tracer of the bolometric luminosity evolution. Then, we rescale these average count rates with the estimated distance of each system (see Table \ref{tab1} and references therein for the used distances) to obtain a representative quantity ($X_r$) that is proportional to the X-ray luminosity. Hence, the $X_r$ parameter provides a crude estimation of the average mass accretion rate; thus, the average torque exerted on the pulsar. We also include the noise strength measurement of GX 1+4 at similar timescales, presented in \cite{2017SerimB}.
The resulting diagram is illustrated in Figure \ref{LvN}.

Even with such a crude approach, the noise strengths exhibit a clear correlation with the representative quantity $X_r$ with a Pearson correlation coefficient of 0.95. Unfortunately, the wind and disk accretion schemes are indistinguishable due to the low number of sample sources, even though wind-fed sources Vela X-1 and 4U 1538-52 reside at the lower part of the diagram due to their low noise levels. Similarly, \cite{1993Baykal} examined the noise behavior of several accreting systems and suggested that the X-ray luminosities correlate with the timing noise in such systems. It should be noted that previous studies suggest that there is no correlation between X-ray luminosity and timing noise amplitudes in isolated sources (e.g. \cite{2019CerriSerim,1985Cordes}). The mass accretion rate, exerted torque, and X-ray luminosity are linked in accreting systems. Higher mass accretion rates can accompany larger inhomogeneities in the accretion flow which should yield higher torque fluctuations. Thus, the accretion process should impact the measured timing noise strengths. Hence, $P_{\Delta\dot{\nu}}$ - $X_r$ correlation should indeed signify the presence of accretion in the system. For instance, the amplitude of the external torque indicated in Equation \ref{torqeq} is related to the mass accretion rate ($\dot{M}$) as \citep{1979Ghosh}:\begin{equation}
N_0 = \dot{M} \sqrt{G M_x R_m}
\end{equation}
where $M_x$ and $R_m$ represent the mass and magnetospheric radius of the neutron star, respectively. On the other hand, $R_m$ also conveys $\dot{M}$ dependency. In the \cite{1979Ghosh} picture, this relation can be expressed as:
\begin{equation}
R_m = 0.52\, R_A \, \mu^{4/7} (2GM_X)^{-1/7} \dot{M}^{-2/7}
\end{equation}
where $R_A$ is the Alven radius and $\mu$ is the dipole moment of the pulsar. Hence, the torque relation becomes:
\begin{equation}
N_0 = 0.49 \,(GM_X )^{3/7} \mu^{2/7} \dot{M}^{6/7}.
\end{equation}
In the shot-noise model described in Section \ref{sec4}, amplitudes of the power density estimates, $P_{\Delta\dot{\nu}}$, are proportional to $\delta L^2$ (inherently from $N_0^{2}$). Therefore, the noise strengths should scale with $\sim\mu^{4/7}\dot{M}^{12/7}$ in the $P_{\Delta\dot{\nu}}$ vs $X_r$ diagram shown in Figure \ref{LvN}. However, a simple power law fit for $P_{\Delta\dot{\nu}}$ vs $X_r$ diagram yield an exponent of $5.9\pm2.4$ (90\% confidence level), which indicates that the luminosity (thus $\dot{M}$) dependence of $P_{\Delta\dot{\nu}}$ may be much stronger than that is anticipated in Ghosh \& Lamb model.  Each individual pulsar in our sample possesses a different magnetic field strength; however, they are all in similar order ($ 2.1- 3.6 \times 10^{12}$ G). Therefore, $\mu^{4/7}$ dependency can not alter the exponent significantly but may only lead to scattering around the mean value. Thus, it can not be the sole reason for this discrepancy. This discrepancy may arise from several factors. First of all, we have a limited number of sample sources, each of which exhibits different timing noise behavior as discussed in Section \ref{sources}. To be more specific, they are composed of different accretion schemes (e.g. wind-fed: Vela X-1 and 4U 1538-52, disk-fed: Her X-1 and Cen X-3) which leads to diverse power density spectrum characteristics whose exponents range from $-3$ to $0$. Since their noise strengths are frequency dependent, comparison at a certain timescale (in our case 1000 days) might affect the results. For example, a comparison at a smaller timescale would decrease the exponent. On the other hand, it should be noted that the \cite{1979Ghosh} derivations consider the case where the pulsar's magnetic and spin axes are aligned. The Alfven radius is significantly affected in the case of oblique rotators \citep{1997Wang,2018Bozzo} which substantially modify the scaling factors of torque fluctuations if the obliquenesses of the pulsars vary. In addition, there is a considerable deviation in the distance estimates of the pulsars when different methods are used. For example,  \cite{2018Treuz} examined a sample of galactic X-ray pulsars and compare the distance estimates achieved via various methods, including \emph{Gaia} distance estimates via Bayesian probabilistic analysis \citep{2018BailerJones} and parallax inversion. They noticed that there is a systematic overestimation of pulsar distances (for $> 5$ kpc) obtained with conventional methods when they are compared with \emph{Gaia} distances. Such systematic differences would have a direct impact on the slope of the $P_{\Delta\dot{\nu}}$-$X_r$ diagram. Nevertheless, it is remarkable that a $P_{\Delta\dot{\nu}}$ vs $X_r$ correlation is still observable under such circumstances.

\section{Discussion and Conclusion}
In this study, we examine timing noise properties of seven persistent accreting systems monitored by \emph{Fermi}/GBM. Using the existing frequency measurements, we model the regular rotation of the pulsars in the system with a quadratic trend and presume the residuals of the model as noise fluctuations. We first explore the time-dependent noise structure of each source and then compare the results with isolated pulsars.

\cite{1997Bildsten} examined the time-dependent timing noise nature of these objects using \emph{CGRO}/BATSE data.
They reported that the power spectra of wind-fed sources 4U 1538-52, Vela X-1, and GX 301-2 accord with a white torque noise structure.
On the other hand, their results on LMXB systems Her X-1 and 4U 1626-67, which are expected to have an accretion disk, possess $\omega^{-2}$ dependence in their power spectra of pulse frequency derivative fluctuations.
Furthermore, GX 1+4 and OAO 1657-415 are shown to exhibit flicker noise \citep{2017SerimB,1997Bildsten}. Due to the limited time span of the input frequency series, \cite{1997Bildsten} could not provide definite conclusions for the power spectral shape of the sources. In the light of new frequency measurements provided by \emph{Fermi}/GBM, we expand the timing noise analysis timescales up to ${\sim}10000$ days.

In the earlier studies, the models to describe the power spectral continuum of pulsars concentrate on the possible inner response to the external torques \citep{1991Baykal, 1997Baykal, 1986Alpar, 2017Gugercinoglu, 2021Meyers1,2021Meyers2} to reveal information about the interior structure of neutron stars. The timing noise contribution of neutron star interior is expected to arise on crust$-$core coupling timescales ${\sim}10{-}100$  P$_{\rm{pulse}}$ \citep{1984Alpar,2009Alpar}.  It should be noted that the vortex creep can also induce timing noise which can be effective timescales of several months \citep{2017Gugercinoglu}. However, in our power density spectra, the crust$-$core coupling timescales are dominated by instrumental noise, and the contribution of vortex creep is expected to modify the noise spectrum in the order of $I_s/I$ \citep{2017Gugercinoglu} which make such contributions indistinguishable due to the resolution of the power spectra. Here, we focus on the long-term properties of the power density spectra, thus we consider noise processes solely originating from external torques, neglecting the interior response of the neutron star with the aim of probing the nature of the accretion flow in those systems.

The examined sources exhibit various characteristics of timing noise, implying that the origins of their noise processes might differ. Our results are consistent with that of \cite{1997Bildsten} within their relatively narrow analysis frequency range. However, we provide a more detailed view of the noise behavior of these sources, especially on long timescales. Known wind-fed systems Vela X-1 and 4U 1538-52 show clear white noise formations extending over very long timescales. The power density spectra of OAO 1657-415 and GX 301-2 reveal flicker noise components that are possibly saturated on longer timescales (i.e. $\tau_2> 10^8$ s).
Interestingly, GX 301-2 also exhibits a white noise component up to ${\sim}400$ day timescales. In both systems, we cannot rule out the possible red noise trend extending beyond that timescale. It is possible that these systems may form enduring flicker noise components similar to the case of GX 1+4 \citep{2017SerimB}. The power spectrum of 4U 1626-67 is characterized by a strong red noise component towards low analysis frequencies, which possibly surpasses our analysis frequency range. On the high analysis frequencies, the power density estimates of 4U 1626-67 rapidly fall below the instrumental noise level, possibly owing to the ultra-compact binary nature of the system.  Her X-1 and Cen X-3 seem to possess $\omega^{-2}$ type red noise component in their power spectra which are saturated over long timescales.

We observe that the known disk-fed sources are characterized by a $1/\omega^2$ type red noise structure which evolves into a white noise component at longer timescales, which resembles to the case of 2S 1417-624 \citep{2022Serim}. The transient BeXRB 2S 1417-624 exhibits the same kind of structure and it has been argued that the saturation of the timing noise strength at long timescales is possibly owed to accretion from a cold depleted disk \citep{2022Serim} where the material enters to the magnetosphere of the pulsar via plasma instabilities. They further show that the red noise component can be eliminated using the $\dot{\nu} \propto L$ correlation, revealing that this noise structure is indeed associated with accretion. The noise saturation of the transient system 2S 1417-624  was observed around ${\sim}100$ day timescale \citep{2022Serim}, whereas the power density spectrum of Her X-1 and Cen X-3 saturates at longer timescales (of a factor ${\sim}4$), which may indicate that the accretion disk in these persistent systems is rather more enduring. Moreover, the observed light curves of such sources\footnote{https://gammaray.nsstc.nasa.gov/gbm/science/pulsars.html} are consistent with the assumption of monochromatic torque events for noise process from an accretion disk. Similarly, the saturation of the timing noise amplitude at long timescales of Her X-1 and Cen X-3 possibly hints at the presence of a cold-depleted disk at these timescales. 

On the other hand, the interpretation of the flicker-type noise structure is not straightforward. The generation of such noise structures implies that there should be the contribution of more than one process or the contribution of a single process with varying timescales. When examined carefully, the light curve of the sources, which generate a flicker-type structure in their power spectra, can be depicted with luminosity variations with different durations. The pulse frequency fluctuations in accreting pulsars are associated with their mass accretion rate and hence with their luminosity. Therefore, luminosity variations suggest the existence of a colored noise process as described in Section \ref{sec4}, which perhaps indicates that the fluctuations at outer parts of the disk (when it is present) influence the fluctuations at inner radii, if the variation timescales are similar or larger than the viscous timescales at these radii \citep{1997Lyubarskii}.
The timescales obtained from modeling are larger than the viscous timescales expected in such systems \citep{2011Icdem}. Thus, the flicker noise component in the power spectra of the pulse frequency derivatives may hint at variations at different radii of the accretion disk. Thus, such a noise component can be triggered by a transient accretion disk \citep{2017SerimB} that exhibits independent viscous fluctuations at different radii in the accretion disk. It should be noted that both OAO 1657-415 and GX 301-2 are suggested to be wind-fed sources with occasional accretion disk formation \citep{2012Jenke,1997Koh,2019Nabizadeh}. The observed flicker noise component may also arise due to the co-existence of two accretion regimes or the cyclic transition between them. 

The case of  4U 1626-67 is rather intriguing. The power density spectrum of 4U 1626-67 shows a steeper red noise component with $\propto1/\omega^3$ when compared with the disk-fed sources Her X-1 and Cen X-3.  The presence of such a steep red noise component is observed in magnetars and attributed to a magnetic origin \citep{2019CerriSerim}. However, the red noise component observed in magnetars arises at shorter timescales. This source is a well-known ultra-compact system, containing a very low mass companion \citep{1971Giacconi, 1977McClintock}. Moreover, it is the only ultra-compact system that hosts a strongly magnetized pulsar \citep{2019Schulz}. The unique case of 4U 1626-67 suggests that there should be different factors to consider for the torque noise processes in accreting systems besides the presence of wind/disk accretion and their magnetic field strengths, such as variations of the inner disk radius \citep{2022Gencali}. More observations are required to reveal the saturation timescales of the noise process (if there are any) and establish a proper model for its power density spectrum.

Apart from the noise spectral characteristics of individual sources, we examine the correlation of noise strengths at similar timescales with physical parameters. Magnetars are shown to possess a correlation between $B$-fields and timing noise strengths \citep{2013Tsang,2019CerriSerim} whereas the radio pulsars seem to have a weaker correlation of this type \citep{2010Hobbs}.  We find accreting pulsars in this study are not correlated with the $B$-fields derived from CRSFs. Furthermore, they reside at a distinct location from the magnetar group on the $B$-field vs. $S_r$ diagram. Although the number of samples in our investigation is insufficient to draw firm conclusions, the absence of such correlation is anticipated since the timing noise of these sources should be dominated by external torques due to accretion, not by magnetospheric variations (except for the possibility for 4U 1626-67). Similar conclusions may apply to magnetars as well.
The accretion from a fossil disk should not dominate the torque fluctuations of magnetars. Therefore, we further examine the connection between X-ray luminosity and noise strengths which is shown to be present in accreting pulsars \citep{1993Baykal}. We seek such correlation using the \emph{Swift}/BAT count rates that are scaled up with the source distances. Despite the limited energy range (15$-$50 keV) of the count rate measurements, we observe that the representative quantity $X_r$ is correlated with the measured noise strengths. Similar to the results of \cite{1993Baykal}, there is no clear distinction between wind-fed and disk-fed sources in the $X_r$ vs  $P_{\Delta\dot{\nu}}$ diagram; however, the timing noise levels measured for wind-fed systems Vela X-1 and 4U 1538$-$52 are relatively low when compared with disk-fed sources. The relation between $P_{\Delta\dot{\nu}}$ -$X_r$ at 1000-day timescale reveals a stronger luminosity dependence than predicted by Ghosh \& Lamb model. As the magnetic field strength distribution of our pulsar sample is very narrow, we do not expect it to have significant effects on $P_{\Delta\dot{\nu}}$ -$X_r$ relation. However, the strong luminosity dependence of the noise strengths may imply oversimplification in the noise strength comparison, considering the unknown obliquenesses, distance uncertainties, and varying accretion schemes. Despite the mentioned caveats, the fact that a correlation between $P_{\Delta\dot{\nu}}$ -$X_r$ is still apparent is noteworthy.

\section*{Acknowledgements}

We acknowledge the support from T\"{U}B\.{I}TAK, (The Scientific and Technological Research Council of Turkey) through the research project MFAG 118F037. The authors thank Prof. Dr. Sıtkı Çağdaş İnam and Çağatay Kerem Dönmez for their valuable remarks that assisted in the improvement of this manuscript.  We would like to thank the anonymous referee for the insightful comments that made it easier to improve the manuscript.

\section*{Data Availability}
All the input data sets (i.e. \emph{CGRO}/BATSE and \emph{Fermi}/GBM frequency measurements and \emph{Swift}/BAT count rates) used in this study are publicly available through their websites indicated in the main text.




\begin{thebibliography}{}
\makeatletter
\relax
\def\mn@urlcharsother{\let\do\@makeother \do\$\do\&\do\#\do\^\do\_\do\%\do\~}
\def\mn@doi{\begingroup\mn@urlcharsother \@ifnextchar [ {\mn@doi@}
  {\mn@doi@[]}}
\def\mn@doi@[#1]#2{\def\@tempa{#1}\ifx\@tempa\@empty \href
  {http://dx.doi.org/#2} {doi:#2}\else \href {http://dx.doi.org/#2} {#1}\fi
  \endgroup}
\def\mn@eprint#1#2{\mn@eprint@#1:#2::\@nil}
\def\mn@eprint@arXiv#1{\href {http://arxiv.org/abs/#1} {{\tt arXiv:#1}}}
\def\mn@eprint@dblp#1{\href {http://dblp.uni-trier.de/rec/bibtex/#1.xml}
  {dblp:#1}}
\def\mn@eprint@#1:#2:#3:#4\@nil{\def\@tempa {#1}\def\@tempb {#2}\def\@tempc
  {#3}\ifx \@tempc \@empty \let \@tempc \@tempb \let \@tempb \@tempa \fi \ifx
  \@tempb \@empty \def\@tempb {arXiv}\fi \@ifundefined
  {mn@eprint@\@tempb}{\@tempb:\@tempc}{\expandafter \expandafter \csname
  mn@eprint@\@tempb\endcsname \expandafter{\@tempc}}}

\bibitem[\protect\citeauthoryear{{Abbott} et~al.,}{{Abbott}
  et~al.}{2017}]{2017Abbott}
{Abbott} B.~P.,  et~al., 2017, \mn@doi [\prl] {10.1103/PhysRevLett.119.161101},
  \href {https://ui.adsabs.harvard.edu/abs/2017PhRvL.119p1101A} {119, 161101}

\bibitem[\protect\citeauthoryear{{Alpar}, {Langer}  \& {Sauls}}{{Alpar}
  et~al.}{1984}]{1984Alpar}
{Alpar} M.~A.,  {Langer} S.~A.,   {Sauls} J.~A.,  1984, \mn@doi [\apj]
  {10.1086/162232}, \href
  {https://ui.adsabs.harvard.edu/abs/1984ApJ...282..533A} {282, 533}

\bibitem[\protect\citeauthoryear{{Alpar}, {Nandkumar}  \& {Pines}}{{Alpar}
  et~al.}{1986}]{1986Alpar}
{Alpar} M.~A.,  {Nandkumar} R.,   {Pines} D.,  1986, \mn@doi [\apj]
  {10.1086/164765}, \href
  {https://ui.adsabs.harvard.edu/abs/1986ApJ...311..197A} {311, 197}

\bibitem[\protect\citeauthoryear{{Arzoumanian}, {Nice}, {Taylor}  \&
  {Thorsett}}{{Arzoumanian} et~al.}{1994}]{1994Arzoumanian}
{Arzoumanian} Z.,  {Nice} D.~J.,  {Taylor} J.~H.,   {Thorsett} S.~E.,  1994,
  \mn@doi [ApJ] {10.1086/173760}, \href
  {https://ui.adsabs.harvard.edu/abs/1994ApJ...422..671A} {422, 671}

\bibitem[\protect\citeauthoryear{{Bahcall} \& {Bahcall}}{{Bahcall} \&
  {Bahcall}}{1972}]{1972Bahcall}
{Bahcall} J.~N.,  {Bahcall} N.~A.,  1972, \mn@doi [\apjl] {10.1086/181070},
  \href {https://ui.adsabs.harvard.edu/abs/1972ApJ...178L...1B} {178, L1}

\bibitem[\protect\citeauthoryear{{Bailer-Jones}, {Rybizki}, {Fouesneau},
  {Mantelet}  \& {Andrae}}{{Bailer-Jones} et~al.}{2018}]{2018BailerJones}
{Bailer-Jones} C.~A.~L.,  {Rybizki} J.,  {Fouesneau} M.,  {Mantelet} G.,
  {Andrae} R.,  2018, \mn@doi [\aj] {10.3847/1538-3881/aacb21}, \href
  {https://ui.adsabs.harvard.edu/abs/2018AJ....156...58B} {156, 58}

\bibitem[\protect\citeauthoryear{{Baykal}}{{Baykal}}{1997}]{1997Baykal}
{Baykal} A.,  1997, \aap, \href
  {https://ui.adsabs.harvard.edu/abs/1997A&A...319..515B} {319, 515}

\bibitem[\protect\citeauthoryear{{Baykal}}{{Baykal}}{2000}]{2000Baykal}
{Baykal} A.,  2000, \mn@doi [\mnras] {10.1046/j.1365-8711.2000.03249.x}, \href
  {https://ui.adsabs.harvard.edu/abs/2000MNRAS.313..637B} {313, 637}

\bibitem[\protect\citeauthoryear{{Baykal} \& {Ögelman}}{{Baykal} \&
  {Ögelman}}{1993}]{1993Baykal}
{Baykal} A.,  {Ögelman} H.,  1993, \aap, \href
  {https://ui.adsabs.harvard.edu/abs/1993A&A...267..119B} {267, 119}

\bibitem[\protect\citeauthoryear{{Baykal}, {Alpar}  \& {Kiziloglu}}{{Baykal}
  et~al.}{1991}]{1991Baykal}
{Baykal} A.,  {Alpar} A.,   {Kiziloglu} U.,  1991, \aap, \href
  {https://ui.adsabs.harvard.edu/abs/1991A&A...252..664B} {252, 664}

\bibitem[\protect\citeauthoryear{{Baykal}, {Inam}  \& {Beklen}}{{Baykal}
  et~al.}{2006}]{2006Baykal}
{Baykal} A.,  {Inam} S.~{\c{C}}.,   {Beklen} E.,  2006, \mn@doi [AAP]
  {10.1051/0004-6361:20054616}, \href
  {https://ui.adsabs.harvard.edu/abs/2006A&A...453.1037B} {453, 1037}

\bibitem[\protect\citeauthoryear{{Bildsten} et~al.,}{{Bildsten}
  et~al.}{1997}]{1997Bildsten}
{Bildsten} L.,  et~al., 1997, \mn@doi [ApJS] {10.1086/313060}, \href
  {https://ui.adsabs.harvard.edu/abs/1997ApJS..113..367B} {113, 367}

\bibitem[\protect\citeauthoryear{{Boynton}, {Groth}, {Hutchinson}, {Nanos},
  {Partridge}  \& {Wilkinson}}{{Boynton} et~al.}{1972}]{1972Boynton}
{Boynton} P.~E.,  {Groth} E.~J.,  {Hutchinson} D.~P.,  {Nanos} G.~P. J.,
  {Partridge} R.~B.,   {Wilkinson} D.~T.,  1972, \mn@doi [\apj]
  {10.1086/151550}, \href{https://ui.adsabs.harvard.edu/abs/1972ApJ...175..217B} {175, 217}
  
\bibitem[\protect\citeauthoryear{Bozzo et al.}{2018}]{2018Bozzo}
 Bozzo E., Ascenzi S., Ducci L., Papitto A., Burderi L., Stella L., 2018, \mn@doi [\aap] {10.1051/0004-6361/201732004}, \href{https://ui.adsabs.harvard.edu/abs/2018A&A...617A.126B}{617, A126}:
\bibitem[\protect\citeauthoryear{{Burderi}, {Robba}, {La Barbera}  \&
  {Guainazzi}}{{Burderi} et~al.}{1997}]{1997Burderi}
{Burderi} L.,  {Robba} N.~R.,  {La Barbera} N.,   {Guainazzi} M.,  1997,
  \mn@doi [\apj] {10.1086/304071}, \href
  {https://ui.adsabs.harvard.edu/abs/1997ApJ...481..943B} {481, 943}

\bibitem[\protect\citeauthoryear{{Camero-Arranz}, {Finger}, {Ikhsanov},
  {Wilson-Hodge}  \& {Beklen}}{{Camero-Arranz} et~al.}{2010}]{2010CameroArranz}
{Camero-Arranz} A.,  {Finger} M.~H.,  {Ikhsanov} N.~R.,  {Wilson-Hodge} C.~A.,
   {Beklen} E.,  2010, \mn@doi [ApJ] {10.1088/0004-637X/708/2/1500}, \href
  {https://ui.adsabs.harvard.edu/abs/2010ApJ...708.1500C} {708, 1500}

\bibitem[\protect\citeauthoryear{{Chakrabarty}}{{Chakrabarty}}{1998}]{1998Chakrabarty}
{Chakrabarty} D.,  1998, \mn@doi [ApJ] {10.1086/305035}, \href
  {https://ui.adsabs.harvard.edu/abs/1998ApJ...492..342C} {492, 342}

\bibitem[\protect\citeauthoryear{{Chakrabarty} et~al.,}{{Chakrabarty}
  et~al.}{1993}]{1993Chakrabarty}
{Chakrabarty} D.,  et~al., 1993, \mn@doi [\apjl] {10.1086/186715}, \href
  {https://ui.adsabs.harvard.edu/abs/1993ApJ...403L..33C} {403, L33}

\bibitem[\protect\citeauthoryear{{Chakrabarty} et~al.,}{{Chakrabarty}
  et~al.}{1997}]{1997Chakrabarty}
{Chakrabarty} D.,  et~al., 1997, \mn@doi [ApJ] {10.1086/303445}, \href
  {https://ui.adsabs.harvard.edu/abs/1997ApJ...474..414C} {474, 414}

\bibitem[\protect\citeauthoryear{{Chandra}, {Roy}, {Agrawal}  \&
  {Choudhury}}{{Chandra} et~al.}{2021}]{2021Chandra}
{Chandra} A.~D.,  {Roy} J.,  {Agrawal} P.~C.,   {Choudhury} M.,  2021, \mn@doi
  [\mnras] {10.1093/mnras/stab2382}, \href
  {https://ui.adsabs.harvard.edu/abs/2021MNRAS.508.4429C} {508, 4429}

\bibitem[\protect\citeauthoryear{{Clark}, {Woo}, {Nagase}, {Makishima}  \&
  {Sakao}}{{Clark} et~al.}{1990}]{1990Clark}
{Clark} G.~W.,  {Woo} J.~W.,  {Nagase} F.,  {Makishima} K.,   {Sakao} T.,
  1990, \mn@doi [\apj] {10.1086/168614}, \href
  {https://ui.adsabs.harvard.edu/abs/1990ApJ...353..274C} {353, 274}

\bibitem[\protect\citeauthoryear{{Corbet}, {Coley}, {Krimm}, {Pottschmidt}  \&
  {Roche}}{{Corbet} et~al.}{2021}]{2021Corbet}
{Corbet} R. H.~D.,  {Coley} J.~B.,  {Krimm} H.~A.,  {Pottschmidt} K.,   {Roche}
  P.,  2021, \mn@doi [\apj] {10.3847/1538-4357/abc477}, \href
  {https://ui.adsabs.harvard.edu/abs/2021ApJ...906...13C} {906, 13}

\bibitem[\protect\citeauthoryear{{Cordes}}{{Cordes}}{1980}]{1980Cordes2}
{Cordes} J.~M.,  1980, \mn@doi [\apj] {10.1086/157861}, \href
  {https://ui.adsabs.harvard.edu/abs/1980ApJ...237..216C} {237, 216}

\bibitem[\protect\citeauthoryear{{Cordes} \& {Downs}}{{Cordes} \&
  {Downs}}{1985}]{1985Cordes}
{Cordes} J.~M.,  {Downs} G.~S.,  1985, \mn@doi [\apjs] {10.1086/191076}, \href
  {https://ui.adsabs.harvard.edu/abs/1985ApJS...59..343C} {59, 343}

\bibitem[\protect\citeauthoryear{{Cordes} \& {Helfand}}{{Cordes} \&
  {Helfand}}{1980}]{1980Cordes1}
{Cordes} J.~M.,  {Helfand} D.~J.,  1980, \mn@doi [\apj] {10.1086/158150}, \href
  {https://ui.adsabs.harvard.edu/abs/1980ApJ...239..640C} {239, 640}
  
\bibitem[\protect\citeauthoryear{{{\c{C}}erri-Serim}, {Serim}, {{\c{S}}ahiner},
  {Inam}  \& {Baykal}}{{{\c{C}}erri-Serim} et~al.}{2019}]{2019CerriSerim}
{{\c{C}}erri-Serim} D.,  {Serim} M.~M.,  {{\c{S}}ahiner} {\c{S}}.,  {Inam}
  S.~{\c{C}}.,   {Baykal} A.,  2019, \mn@doi [\mnras] {10.1093/mnras/sty3213},
  \href {https://ui.adsabs.harvard.edu/abs/2019MNRAS.485....2C} {485, 2}
  
\bibitem[\protect\citeauthoryear{{Davison}, {Watson}  \& {Pye}}{{Davison}
  et~al.}{1977}]{1977Davison}
{Davison} P.~J.~N.,  {Watson} M.~G.,   {Pye} J.~P.,  1977, \mn@doi [MNRAS]
  {10.1093/mnras/181.1.73P}, \href
  {https://ui.adsabs.harvard.edu/abs/1977MNRAS.181P..73D} {181, 73}

\bibitem[\protect\citeauthoryear{{Deeter}}{{Deeter}}{1984}]{1984Deeter}
{Deeter} J.~E.,  1984, \mn@doi [ApJ] {10.1086/162122}, \href
  {https://ui.adsabs.harvard.edu/abs/1984ApJ...281..482D} {281, 482}

\bibitem[\protect\citeauthoryear{{Deeter}, {Boynton}, {Lamb}  \&
  {Zylstra}}{{Deeter} et~al.}{1989}]{1989Deeter}
{Deeter} J.~E.,  {Boynton} P.~E.,  {Lamb} F.~K.,   {Zylstra} G.,  1989, \mn@doi
  [ApJ] {10.1086/167017}, \href
  {https://ui.adsabs.harvard.edu/abs/1989ApJ...336..376D} {336, 376}

\bibitem[\protect\citeauthoryear{{Doroshenko}, {Santangelo}, {Suleimanov},
  {Kreykenbohm}, {Staubert}, {Ferrigno}  \& {Klochkov}}{{Doroshenko}
  et~al.}{2010}]{2010Doroshenko}
{Doroshenko} V.,  {Santangelo} A.,  {Suleimanov} V.,  {Kreykenbohm} I.,
  {Staubert} R.,  {Ferrigno} C.,   {Klochkov} D.,  2010, \mn@doi [\aap]
  {10.1051/0004-6361/200912951}, \href
  {https://ui.adsabs.harvard.edu/abs/2010A&A...515A..10D} {515, A10}

\bibitem[\protect\citeauthoryear{{Ertan} \& {{\c{C}}al{\i}{\c{s}}kan}}{{Ertan}
  \& {{\c{C}}al{\i}{\c{s}}kan}}{2006}]{2006Ertan}
{Ertan} {\"U}.,  {{\c{C}}al{\i}{\c{s}}kan} {\c{S}}.,  2006, \mn@doi [\apjl]
  {10.1086/508347}, \href
  {https://ui.adsabs.harvard.edu/abs/2006ApJ...649L..87E} {649, L87}

\bibitem[\protect\citeauthoryear{{F{\"u}rst} et~al.,}{{F{\"u}rst}
  et~al.}{2014}]{2014bFurst}
{F{\"u}rst} F.,  et~al., 2014, \mn@doi [\apj] {10.1088/0004-637X/780/2/133},
  \href {https://ui.adsabs.harvard.edu/abs/2014ApJ...780..133F} {780, 133}

\bibitem[\protect\citeauthoryear{{Gen{\c{c}}ali} et~al.,}{{Gen{\c{c}}ali}
  et~al.}{2022}]{2022Gencali}
{Gen{\c{c}}ali} A.~A.,  et~al., 2022, \mn@doi [\aap]
  {10.1051/0004-6361/202141772}, \href
  {https://ui.adsabs.harvard.edu/abs/2022A&A...658A..13G} {658, A13}
  
\bibitem[\protect\citeauthoryear{Ghosh \& Lamb}{1979}]{1979Ghosh} Ghosh P., Lamb F.~K., 1979, \mn@doi [\apj] {10.1086/157285},  \href{https://ui.adsabs.harvard.edu/abs/1979ApJ...232..259G}{232, 259}

\bibitem[\protect\citeauthoryear{{Giacconi}, {Gursky}, {Kellogg}, {Schreier}
  \& {Tananbaum}}{{Giacconi} et~al.}{1971}]{1971Giacconi}
{Giacconi} R.,  {Gursky} H.,  {Kellogg} E.,  {Schreier} E.,   {Tananbaum} H.,
  1971, \mn@doi [ApJL] {10.1086/180762}, \href
  {https://ui.adsabs.harvard.edu/abs/1971ApJ...167L..67G} {167, L67}

\bibitem[\protect\citeauthoryear{{Giacconi}, {Gursky}, {Kellogg}, {Levinson},
  {Schreier}  \& {Tananbaum}}{{Giacconi} et~al.}{1973}]{1973Giacconi}
{Giacconi} R.,  {Gursky} H.,  {Kellogg} E.,  {Levinson} R.,  {Schreier} E.,
  {Tananbaum} H.,  1973, \mn@doi [\apj] {10.1086/152321}, \href
  {https://ui.adsabs.harvard.edu/abs/1973ApJ...184..227G} {184, 227}

\bibitem[\protect\citeauthoryear{{Goncharov}, {Zhu}  \& {Thrane}}{{Goncharov}
  et~al.}{2020}]{2020Goncharov}
{Goncharov} B.,  {Zhu} X.-J.,   {Thrane} E.,  2020, \mn@doi [\mnras]
  {10.1093/mnras/staa2081}, \href
  {https://ui.adsabs.harvard.edu/abs/2020MNRAS.497.3264G} {497, 3264}

\bibitem[\protect\citeauthoryear{{Groth}}{{Groth}}{1975}]{1975Groth}
{Groth} E.~J.,  1975, \mn@doi [\apjs] {10.1086/190354}, \href
  {https://ui.adsabs.harvard.edu/abs/1975ApJS...29..453G} {29, 453}

\bibitem[\protect\citeauthoryear{{G{\"u}gercino{\v{g}}lu} \&
  {Alpar}}{{G{\"u}gercino{\v{g}}lu} \& {Alpar}}{2017}]{2017Gugercinoglu}
{G{\"u}gercino{\v{g}}lu} E.,  {Alpar} M.~A.,  2017, \mn@doi [\mnras]
  {10.1093/mnras/stx1937}, \href
  {https://ui.adsabs.harvard.edu/abs/2017MNRAS.471.4827G} {471, 4827}

\bibitem[\protect\citeauthoryear{{Hemphill}, {Rothschild}, {Markowitz},
  {F{\"u}rst}, {Pottschmidt}  \& {Wilms}}{{Hemphill}
  et~al.}{2014}]{2014Hemphill}
{Hemphill} P.~B.,  {Rothschild} R.~E.,  {Markowitz} A.,  {F{\"u}rst} F.,
  {Pottschmidt} K.,   {Wilms} J.,  2014, \mn@doi [\apj]
  {10.1088/0004-637X/792/1/14}, \href
  {https://ui.adsabs.harvard.edu/abs/2014ApJ...792...14H} {792, 14}

\bibitem[\protect\citeauthoryear{{Hiltner}, {Werner}  \& {Osmer}}{{Hiltner}
  et~al.}{1972}]{1972Hiltner}
{Hiltner} W.~A.,  {Werner} J.,   {Osmer} P.,  1972, \mn@doi [\apjl]
  {10.1086/180976}, \href
  {https://ui.adsabs.harvard.edu/abs/1972ApJ...175L..19H} {175, L19}

\bibitem[\protect\citeauthoryear{{Hobbs}, {Lyne}  \& {Kramer}}{{Hobbs}
  et~al.}{2010}]{2010Hobbs}
{Hobbs} G.,  {Lyne} A.~G.,   {Kramer} M.,  2010, \mn@doi [\mnras]
  {10.1111/j.1365-2966.2009.15938.x}, \href
  {https://ui.adsabs.harvard.edu/abs/2010MNRAS.402.1027H} {402, 1027}

\bibitem[\protect\citeauthoryear{{Hulleman}, {Tennant}, {van Kerkwijk},
  {Kulkarni}, {Kouveliotou}  \& {Patel}}{{Hulleman}
  et~al.}{2001}]{2001Hulleman}
{Hulleman} F.,  {Tennant} A.~F.,  {van Kerkwijk} M.~H.,  {Kulkarni} S.~R.,
  {Kouveliotou} C.,   {Patel} S.~K.,  2001, \mn@doi [\apjl] {10.1086/338478},
  \href {https://ui.adsabs.harvard.edu/abs/2001ApJ...563L..49H} {563, L49}

\bibitem[\protect\citeauthoryear{{I{\c{c}}dem} \& {Baykal}}{{I{\c{c}}dem} \&
  {Baykal}}{2011}]{2011Icdem}
{I{\c{c}}dem} B.,  {Baykal} A.,  2011, \mn@doi [\aap]
  {10.1051/0004-6361/201015810}, \href
  {https://ui.adsabs.harvard.edu/abs/2011A&A...529A...7I} {529, A7}

\bibitem[\protect\citeauthoryear{{Jenke}, {Finger}, {Wilson-Hodge}  \&
  {Camero-Arranz}}{{Jenke} et~al.}{2012}]{2012Jenke}
{Jenke} P.~A.,  {Finger} M.~H.,  {Wilson-Hodge} C.~A.,   {Camero-Arranz} A.,
  2012, \mn@doi [\apj] {10.1088/0004-637X/759/2/124}, \href
  {https://ui.adsabs.harvard.edu/abs/2012ApJ...759..124J} {759, 124}

\bibitem[\protect\citeauthoryear{{Kaper}, {van der Meer}  \& {Najarro}}{{Kaper}
  et~al.}{2006}]{2006Kaper}
{Kaper} L.,  {van der Meer} A.,   {Najarro} F.,  2006, \mn@doi [AAP]
  {10.1051/0004-6361:20065393}, \href
  {https://ui.adsabs.harvard.edu/abs/2006A&A...457..595K} {457, 595}

\bibitem[\protect\citeauthoryear{{Kaplan}, {Kulkarni}, {van Kerkwijk},
  {Rothschild}, {Lingenfelter}, {Marsden}, {Danner}  \& {Murakami}}{{Kaplan}
  et~al.}{2001}]{2001Kaplan}
{Kaplan} D.~L.,  {Kulkarni} S.~R.,  {van Kerkwijk} M.~H.,  {Rothschild} R.~E.,
  {Lingenfelter} R.~L.,  {Marsden} D.,  {Danner} R.,   {Murakami} T.,  2001,
  \mn@doi [\apj] {10.1086/323516}, \href
  {https://ui.adsabs.harvard.edu/abs/2001ApJ...556..399K} {556, 399}

\bibitem[\protect\citeauthoryear{{Kaplan}, {Chakrabarty}, {Wang}  \&
  {Wachter}}{{Kaplan} et~al.}{2009}]{2009Kaplan}
{Kaplan} D.~L.,  {Chakrabarty} D.,  {Wang} Z.,   {Wachter} S.,  2009, \mn@doi
  [\apj] {10.1088/0004-637X/700/1/149}, \href
  {https://ui.adsabs.harvard.edu/abs/2009ApJ...700..149K} {700, 149}

\bibitem[\protect\citeauthoryear{{Kelley}, {Rappaport}, {Clark}  \&
  {Petro}}{{Kelley} et~al.}{1983}]{1983bKelley}
{Kelley} R.~L.,  {Rappaport} S.,  {Clark} G.~W.,   {Petro} L.~D.,  1983,
  \mn@doi [ApJ] {10.1086/161001}, \href
  {https://ui.adsabs.harvard.edu/abs/1983ApJ...268..790K} {268, 790}
  
\bibitem[\protect\citeauthoryear{{van Kerkwijk}, {van Paradijs}, {Zuiderwijk},
  {Hammerschlag-Hensberge}, {Kaper}  \& {Sterken}}{{van Kerkwijk}
  et~al.}{1995}]{1995vanKerkwijk}
{van Kerkwijk} M.~H.,  {van Paradijs} J.,  {Zuiderwijk} E.~J.,
  {Hammerschlag-Hensberge} G.,  {Kaper} L.,   {Sterken} C.,  1995, \aap, \href
  {https://ui.adsabs.harvard.edu/abs/1995A&A...303..483V} {303, 483}
  
\bibitem[\protect\citeauthoryear{{Koh} et~al.,}{{Koh} et~al.}{1997}]{1997Koh}
{Koh} D.~T.,  et~al., 1997, \mn@doi [ApJ] {10.1086/303929}, \href
  {https://ui.adsabs.harvard.edu/abs/1997ApJ...479..933K} {479, 933}

\bibitem[\protect\citeauthoryear{{Kreykenbohm}, {Wilms}, {Coburn}, {Kuster},
  {Rothschild}, {Heindl}, {Kretschmar}  \& {Staubert}}{{Kreykenbohm}
  et~al.}{2004}]{2004Kreykenbohm}
{Kreykenbohm} I.,  {Wilms} J.,  {Coburn} W.,  {Kuster} M.,  {Rothschild} R.~E.,
   {Heindl} W.~A.,  {Kretschmar} P.,   {Staubert} R.,  2004, \mn@doi [\aap]
  {10.1051/0004-6361:20035836}, \href
  {https://ui.adsabs.harvard.edu/abs/2004A&A...427..975K} {427, 975}

\bibitem[\protect\citeauthoryear{{Krzeminski}}{{Krzeminski}}{1974}]{1974Krzeminski}
{Krzeminski} W.,  1974, \mn@doi [ApJL] {10.1086/181609}, \href
  {https://ui.adsabs.harvard.edu/abs/1974ApJ...192L.135K} {192, L135}

\bibitem[\protect\citeauthoryear{{Lower} et~al.,}{{Lower}
  et~al.}{2020}]{2020Lower}
{Lower} M.~E.,  et~al., 2020, \mn@doi [\mnras] {10.1093/mnras/staa615}, \href
  {https://ui.adsabs.harvard.edu/abs/2020MNRAS.494..228L} {494, 228}

\bibitem[\protect\citeauthoryear{{Lyubarskii}}{{Lyubarskii}}{1997}]{1997Lyubarskii}
{Lyubarskii} Y.~E.,  1997, \mn@doi [\mnras] {10.1093/mnras/292.3.679}, \href
  {https://ui.adsabs.harvard.edu/abs/1997MNRAS.292..679L} {292, 679}

\bibitem[\protect\citeauthoryear{{Malacaria}, {Jenke}, {Roberts},
  {Wilson-Hodge}, {Cleveland}, {Mailyan}  \& {GBM Accreting Pulsars Program
  Team}}{{Malacaria} et~al.}{2020}]{2020Malacaria}
{Malacaria} C.,  {Jenke} P.,  {Roberts} O.~J.,  {Wilson-Hodge} C.~A.,
  {Cleveland} W.~H.,  {Mailyan} B.,   {GBM Accreting Pulsars Program Team}
  2020, \mn@doi [ApJ] {10.3847/1538-4357/ab855c}, \href
  {https://ui.adsabs.harvard.edu/abs/2020ApJ...896...90M} {896, 90}

\bibitem[\protect\citeauthoryear{{Mason}, {Clark}, {Norton}, {Negueruela}  \&
  {Roche}}{{Mason} et~al.}{2009}]{2009Mason}
{Mason} A.~B.,  {Clark} J.~S.,  {Norton} A.~J.,  {Negueruela} I.,   {Roche} P.,
   2009, \mn@doi [\aap] {10.1051/0004-6361/200912480}, \href
  {https://ui.adsabs.harvard.edu/abs/2009A&A...505..281M} {505, 281}

\bibitem[\protect\citeauthoryear{{McClintock}, {Bradt}, {Doxsey}, {Jernigan},
  {Canizares}  \& {Hiltner}}{{McClintock} et~al.}{1977}]{1977McClintock}
{McClintock} J.~E.,  {Bradt} H.~V.,  {Doxsey} R.~E.,  {Jernigan} J.~G.,
  {Canizares} C.~R.,   {Hiltner} W.~A.,  1977, \mn@doi [Nature]
  {10.1038/270320a0}, \href
  {https://ui.adsabs.harvard.edu/abs/1977Natur.270..320M} {270, 320}

\bibitem[\protect\citeauthoryear{{Mereghetti}}{{Mereghetti}}{2008}]{2008Mereghetti}
{Mereghetti} S.,  2008, \mn@doi [\aapr] {10.1007/s00159-008-0011-z}, \href
  {https://ui.adsabs.harvard.edu/abs/2008A&ARv..15..225M} {15, 225}

\bibitem[\protect\citeauthoryear{{Meyers}, {Melatos}  \& {O'Neill}}{{Meyers}
  et~al.}{2021a}]{2021Meyers1}
{Meyers} P.~M.,  {Melatos} A.,   {O'Neill} N.~J.,  2021a, \mn@doi [\mnras]
  {10.1093/mnras/stab262}, \href
  {https://ui.adsabs.harvard.edu/abs/2021MNRAS.502.3113M} {502, 3113}

\bibitem[\protect\citeauthoryear{{Meyers}, {O'Neill}, {Melatos}  \&
  {Evans}}{{Meyers} et~al.}{2021b}]{2021Meyers2}
{Meyers} P.~M.,  {O'Neill} N.~J.,  {Melatos} A.,   {Evans} R.~J.,  2021b,
  \mn@doi [\mnras] {10.1093/mnras/stab1952}, \href
  {https://ui.adsabs.harvard.edu/abs/2021MNRAS.506.3349M} {506, 3349}

\bibitem[\protect\citeauthoryear{{Middleditch} \& {Nelson}}{{Middleditch} \&
  {Nelson}}{1976}]{1976Middleditch}
{Middleditch} J.,  {Nelson} J.,  1976, \mn@doi [\apj] {10.1086/154638}, \href
  {https://ui.adsabs.harvard.edu/abs/1976ApJ...208..567M} {208, 567}

\bibitem[\protect\citeauthoryear{{Middleditch}, {Mason}, {Nelson}  \&
  {White}}{{Middleditch} et~al.}{1981}]{1981Middleditch}
{Middleditch} J.,  {Mason} K.~O.,  {Nelson} J.~E.,   {White} N.~E.,  1981,
  \mn@doi [ApJ] {10.1086/158772}, \href
  {https://ui.adsabs.harvard.edu/abs/1981ApJ...244.1001M} {244, 1001}

\bibitem[\protect\citeauthoryear{{Milotti}}{{Milotti}}{2002}]{2002Milotti}
{Milotti} E.,  2002, arXiv e-prints, \href
  {https://ui.adsabs.harvard.edu/abs/2002physics...4033M} {p. physics/0204033}

\bibitem[\protect\citeauthoryear{{Nabizadeh}, {M{\"o}nkk{\"o}nen}, {Tsygankov},
  {Doroshenko}, {Molkov}  \& {Poutanen}}{{Nabizadeh}
  et~al.}{2019}]{2019Nabizadeh}
{Nabizadeh} A.,  {M{\"o}nkk{\"o}nen} J.,  {Tsygankov} S.~S.,  {Doroshenko} V.,
  {Molkov} S.~V.,   {Poutanen} J.,  2019, \mn@doi [AAP]
  {10.1051/0004-6361/201936045}, \href
  {https://ui.adsabs.harvard.edu/abs/2019A&A...629A.101N} {629, A101}

\bibitem[\protect\citeauthoryear{{Nagase}}{{Nagase}}{1989}]{1989Nagase}
{Nagase} F.,  1989, \pasj, \href
  {https://ui.adsabs.harvard.edu/abs/1989PASJ...41....1N} {41, 1}

\bibitem[\protect\citeauthoryear{{Namkham}, {Jaroenjittichai}  \&
  {Johnston}}{{Namkham} et~al.}{2019}]{2019Namkham}
{Namkham} N.,  {Jaroenjittichai} P.,   {Johnston} S.,  2019, \mn@doi [\mnras]
  {10.1093/mnras/stz1671}, \href
  {https://ui.adsabs.harvard.edu/abs/2019MNRAS.487.5854N} {487, 5854}

\bibitem[\protect\citeauthoryear{{Orlandini} et~al.,}{{Orlandini}
  et~al.}{1998}]{1998Orlandini}
{Orlandini} M.,  et~al., 1998, \mn@doi [\apjl] {10.1086/311404}, \href
  {https://ui.adsabs.harvard.edu/abs/1998ApJ...500L.163O} {500, L163}

\bibitem[\protect\citeauthoryear{{{\"O}zel} \& {Freire}}{{{\"O}zel} \&
  {Freire}}{2016}]{2016Ozel}
{{\"O}zel} F.,  {Freire} P.,  2016, \mn@doi [\araa]
  {10.1146/annurev-astro-081915-023322}, \href
  {https://ui.adsabs.harvard.edu/abs/2016ARA&A..54..401O} {54, 401}

\bibitem[\protect\citeauthoryear{{Parkes}, {Murdin}  \& {Mason}}{{Parkes}
  et~al.}{1978}]{1978Parkes}
{Parkes} G.~E.,  {Murdin} P.~G.,   {Mason} K.~O.,  1978, \mn@doi [MNRAS]
  {10.1093/mnras/184.1.73P}, \href
  {https://ui.adsabs.harvard.edu/abs/1978MNRAS.184P..73P} {184, 73P}

\bibitem[\protect\citeauthoryear{{Parkes}, {Mason}, {Murdin}  \&
  {Culhane}}{{Parkes} et~al.}{1980}]{1980Parkes}
{Parkes} G.~E.,  {Mason} K.~O.,  {Murdin} P.~G.,   {Culhane} J.~L.,  1980,
  \mn@doi [\mnras] {10.1093/mnras/191.3.547}, \href
  {https://ui.adsabs.harvard.edu/abs/1980MNRAS.191..547P} {191, 547}

\bibitem[\protect\citeauthoryear{{Parthasarathy} et~al.,}{{Parthasarathy}
  et~al.}{2019}]{2019Parthasarathy}
{Parthasarathy} A.,  et~al., 2019, \mn@doi [\mnras] {10.1093/mnras/stz2383},
  \href {https://ui.adsabs.harvard.edu/abs/2019MNRAS.489.3810P} {489, 3810}

\bibitem[\protect\citeauthoryear{{Parthasarathy} et~al.,}{{Parthasarathy}
  et~al.}{2020}]{2020Parthasarathy}
{Parthasarathy} A.,  et~al., 2020, \mn@doi [\mnras] {10.1093/mnras/staa882},
  \href {https://ui.adsabs.harvard.edu/abs/2020MNRAS.494.2012P} {494, 2012}

\bibitem[\protect\citeauthoryear{{Polidan}, {Pollard}, {Sanford}  \&
  {Locke}}{{Polidan} et~al.}{1978}]{1978Polidan}
{Polidan} R.~S.,  {Pollard} G.~S.~G.,  {Sanford} P.~W.,   {Locke} M.~C.,  1978,
  \mn@doi [\nat] {10.1038/275296a0}, \href
  {https://ui.adsabs.harvard.edu/abs/1978Natur.275..296P} {275, 296}

\bibitem[\protect\citeauthoryear{{Press}}{{Press}}{1978}]{1978Press}
{Press} W.~H.,  1978, Comments on Astrophysics, \href
  {https://ui.adsabs.harvard.edu/abs/1978ComAp...7..103P} {7, 103}

\bibitem[\protect\citeauthoryear{{Revnivtsev} \& {Mereghetti}}{{Revnivtsev} \&
  {Mereghetti}}{2015}]{2015Revnivtsev}
{Revnivtsev} M.,  {Mereghetti} S.,  2015, \mn@doi [\ssr]
  {10.1007/s11214-014-0123-x}, \href
  {https://ui.adsabs.harvard.edu/abs/2015SSRv..191..293R} {191, 293}

\bibitem[\protect\citeauthoryear{{Santangelo}, {del Sordo}, {Segreto}, {dal
  Fiume}, {Orlandini}  \& {Piraino}}{{Santangelo}
  et~al.}{1998}]{1998Santangelo}
{Santangelo} A.,  {del Sordo} S.,  {Segreto} A.,  {dal Fiume} D.,  {Orlandini}
  M.,   {Piraino} S.,  1998, \aap, \href
  {https://ui.adsabs.harvard.edu/abs/1998A&A...340L..55S} {340, L55}

\bibitem[\protect\citeauthoryear{{Schulz}, {Chakrabarty}  \&
  {Marshall}}{{Schulz} et~al.}{2019}]{2019Schulz}
{Schulz} N.~S.,  {Chakrabarty} D.,   {Marshall} H.~L.,  2019, arXiv e-prints,
  \href {https://ui.adsabs.harvard.edu/abs/2019arXiv191111684S} {p.
  arXiv:1911.11684}

\bibitem[\protect\citeauthoryear{{Scott}, {Finger}  \& {Wilson}}{{Scott}
  et~al.}{2003}]{2003Scott}
{Scott} D.~M.,  {Finger} M.~H.,   {Wilson} C.~A.,  2003, \mn@doi [MNRAS]
  {10.1046/j.1365-8711.2003.06825.x}, \href
  {https://ui.adsabs.harvard.edu/abs/2003MNRAS.344..412S} {344, 412}

\bibitem[\protect\citeauthoryear{{Serim}, {{\c{S}}ahiner}, {{\c{C}}erri-Serim},
  {{\.I}nam}  \& {Baykal}}{{Serim} et~al.}{2017a}]{2017SerimB}
{Serim} M.~M.,  {{\c{S}}ahiner} {\c{S}}.,  {{\c{C}}erri-Serim} D.,  {{\.I}nam}
  S.~{\c{C}}.,   {Baykal} A.,  2017a, \mn@doi [\mnras] {10.1093/mnras/stx1045},
  \href {https://ui.adsabs.harvard.edu/abs/2017MNRAS.469.2509S} {469, 2509}

\bibitem[\protect\citeauthoryear{{Serim}, {{\c{S}}ahiner}, {{\c{C}}erri-Serim},
  {Inam}  \& {Baykal}}{{Serim} et~al.}{2017b}]{2017Serim}
{Serim} M.~M.,  {{\c{S}}ahiner} {\c{S}}.,  {{\c{C}}erri-Serim} D.,  {Inam}
  S.~{\c{C}}.,   {Baykal} A.,  2017b, \mn@doi [\mnras] {10.1093/mnras/stx1771},
  \href {https://ui.adsabs.harvard.edu/abs/2017MNRAS.471.4982S} {471, 4982}

\bibitem[\protect\citeauthoryear{{Serim}, {{\"O}z{\"u}do{\u{g}}ru},
  {D{\"o}nmez}, {{\c{S}}ahiner}, {Serim}, {Baykal}  \& {{\.I}nam}}{{Serim}
  et~al.}{2022}]{2022Serim}
{Serim} M.~M.,  {{\"O}z{\"u}do{\u{g}}ru} {\"O}.~C.,  {D{\"o}nmez} {\c{C}}.~K.,
  {{\c{S}}ahiner} {\c{S}}.,  {Serim} D.,  {Baykal} A.,   {{\.I}nam}
  S.~{\c{C}}.,  2022, \mn@doi [\mnras] {10.1093/mnras/stab3547}, \href
  {https://ui.adsabs.harvard.edu/abs/2022MNRAS.510.1438S} {510, 1438}

\bibitem[\protect\citeauthoryear{{Shannon} \& {Cordes}}{{Shannon} \&
  {Cordes}}{2010}]{2010Shannon}
{Shannon} R.~M.,  {Cordes} J.~M.,  2010, \mn@doi [\apj]
  {10.1088/0004-637X/725/2/1607}, \href
  {https://ui.adsabs.harvard.edu/abs/2010ApJ...725.1607S} {725, 1607}

\bibitem[\protect\citeauthoryear{{Sharma}, {Sharma}, {Jain}  \&
  {Dutta}}{{Sharma} et~al.}{2022}]{2022Sharma}
{Sharma} P.,  {Sharma} R.,  {Jain} C.,   {Dutta} A.,  2022, \mn@doi [\mnras]
  {10.1093/mnras/stab3436}, \href
  {https://ui.adsabs.harvard.edu/abs/2022MNRAS.509.5747S} {509, 5747}

\bibitem[\protect\citeauthoryear{{Sidery} \& {Alpar}}{{Sidery} \&
  {Alpar}}{2009}]{2009Alpar}
{Sidery} T.,  {Alpar} M.~A.,  2009, \mn@doi [\mnras]
  {10.1111/j.1365-2966.2009.15575.x}, \href
  {https://ui.adsabs.harvard.edu/abs/2009MNRAS.400.1859S} {400, 1859}

\bibitem[\protect\citeauthoryear{{Stairs}}{{Stairs}}{2003}]{2003Stairs}
{Stairs} I.~H.,  2003, \mn@doi [Living Reviews in Relativity]
  {10.12942/lrr-2003-5}, \href
  {https://ui.adsabs.harvard.edu/abs/2003LRR.....6....5S} {6, 5}

\bibitem[\protect\citeauthoryear{{Staubert}, {Shakura}, {Postnov}, {Wilms},
  {Rothschild}, {Coburn}, {Rodina}  \& {Klochkov}}{{Staubert}
  et~al.}{2007}]{2007Staubert}
{Staubert} R.,  {Shakura} N.~I.,  {Postnov} K.,  {Wilms} J.,  {Rothschild}
  R.~E.,  {Coburn} W.,  {Rodina} L.,   {Klochkov} D.,  2007, \mn@doi [\aap]
  {10.1051/0004-6361:20077098}, \href
  {https://ui.adsabs.harvard.edu/abs/2007A&A...465L..25S} {465, L25}

\bibitem[\protect\citeauthoryear{{Staubert} et~al.,}{{Staubert}
  et~al.}{2019}]{2019Staubert}
{Staubert} R.,  et~al., 2019, \mn@doi [\aap] {10.1051/0004-6361/201834479},
  \href {https://ui.adsabs.harvard.edu/abs/2019A&A...622A..61S} {622, A61}

\bibitem[\protect\citeauthoryear{{Suchy} et~al.,}{{Suchy}
  et~al.}{2008}]{2008Suchy}
{Suchy} S.,  et~al., 2008, \mn@doi [ApJ] {10.1086/527042}, \href
  {https://ui.adsabs.harvard.edu/abs/2008ApJ...675.1487S} {675, 1487}

\bibitem[\protect\citeauthoryear{{Tananbaum}, {Gursky}, {Kellogg}, {Levinson},
  {Schreier}  \& {Giacconi}}{{Tananbaum} et~al.}{1972}]{1972Tananbaum}
{Tananbaum} H.,  {Gursky} H.,  {Kellogg} E.~M.,  {Levinson} R.,  {Schreier} E.,
    {Giacconi} R.,  1972, \mn@doi [ApJL] {10.1086/180968}, \href
  {https://ui.adsabs.harvard.edu/abs/1972ApJ...174L.143T} {174, L143}

\bibitem[\protect\citeauthoryear{Treuz et al.}{2018}]{2018Treuz}
 Treuz S., Doroshenko V., Santangelo A., Staubert R., 2018, arXiv e-prints, \href{https://ui.adsabs.harvard.edu/abs/2018arXiv180611397T}{arXiv:1806.11397}
 
\bibitem[\protect\citeauthoryear{{Tr{\"u}mper}, {Dennerl}, {Kylafis}, {Ertan}
  \& {Zezas}}{{Tr{\"u}mper} et~al.}{2013}]{2013Trumper}
{Tr{\"u}mper} J.~E.,  {Dennerl} K.,  {Kylafis} N.~D.,  {Ertan} {\"U}.,
  {Zezas} A.,  2013, \mn@doi [\apj] {10.1088/0004-637X/764/1/49}, \href
  {https://ui.adsabs.harvard.edu/abs/2013ApJ...764...49T} {764, 49}

\bibitem[\protect\citeauthoryear{{Tsang} \& {Gourgouliatos}}{{Tsang} \&
  {Gourgouliatos}}{2013}]{2013Tsang}
{Tsang} D.,  {Gourgouliatos} K.~N.,  2013, \mn@doi [\apjl]
  {10.1088/2041-8205/773/1/L17}, \href
  {https://ui.adsabs.harvard.edu/abs/2013ApJ...773L..17T} {773, L17}
  
\bibitem[\protect\citeauthoryear{Wang}{1997}]{1997Wang} Wang Y.-M., 1997,  \mn@doi [\apjl] {10.1086/31048},  \href{https://ui.adsabs.harvard.edu/abs/1997ApJ...475L.135W}{475, L135}

\bibitem[\protect\citeauthoryear{{Wang}, {Chakrabarty}  \& {Kaplan}}{{Wang}
  et~al.}{2006}]{2006Wang}
{Wang} Z.,  {Chakrabarty} D.,   {Kaplan} D.~L.,  2006, \mn@doi [\nat]
  {10.1038/nature04669}, \href
  {https://ui.adsabs.harvard.edu/abs/2006Natur.440..772W} {440, 772}

\bibitem[\protect\citeauthoryear{{White} \& {Pravdo}}{{White} \&
  {Pravdo}}{1979}]{1979White}
{White} N.~E.,  {Pravdo} S.~H.,  1979, \mn@doi [\apjl] {10.1086/183089}, \href
  {https://ui.adsabs.harvard.edu/abs/1979ApJ...233L.121W} {233, L121}

\bibitem[\protect\citeauthoryear{{White}, {Mason}, {Huckle}, {Charles}  \&
  {Sanford}}{{White} et~al.}{1976}]{1976White}
{White} N.~E.,  {Mason} K.~O.,  {Huckle} H.~E.,  {Charles} P.~A.,   {Sanford}
  P.~W.,  1976, \mn@doi [\apjl] {10.1086/182281}, \href
  {https://ui.adsabs.harvard.edu/abs/1976ApJ...209L.119W} {209, L119}

\bibitem[\protect\citeauthoryear{{Wilson}, {Fishman}, {Finger}, {Pendleton},
  {Prince}  \& {Chakrabarty}}{{Wilson} et~al.}{1993}]{1993Wilson}
{Wilson} R.~B.,  {Fishman} G.~J.,  {Finger} M.~H.,  {Pendleton} G.~N.,
  {Prince} T.~A.,   {Chakrabarty} D.,  1993, in {Friedlander} M.,  {Gehrels}
  N.,   {Macomb} D.~J.,  eds,  American Institute of Physics Conference Series
  Vol. 280, Compton Gamma-ray Observatory. pp 291--302,
  \mn@doi{10.1063/1.44135}


\makeatother
\end{thebibliography}




\appendix




\bsp	
\label{lastpage}
\end{document}